\documentclass[final,3p,times,sort&compress]{elsarticle}
\usepackage{lipsum}
\usepackage{graphicx}%
\usepackage{amsthm} 
\usepackage{amsmath,amssymb,physics}
\usepackage{hyperref}
\usepackage{url} 
\usepackage{xspace}  %
\usepackage{braket}  %
\usepackage{wrapfig}  %
\usepackage{courier}  %
\usepackage{xcolor} 
\usepackage{comment} %
\usepackage{array} %
\usepackage{%
            soul}
\usepackage{longtable}
\usepackage{xparse}
\usepackage{subfigure}
\usepackage{ulem} %

 \usepackage{setspace}

\newcommand{\Fantomas}{{\sf Fant\^omas}\xspace}  %
\newcommand{\meta}{metamorph\xspace}
\newcommand{\Bezier}{B\'ezier\xspace}
\newcommand{\xFitter}{\mbox{\textsf{xFitter}}\xspace}  %
\newcommand{\FantoPDF}{{\mbox{\sf FantoPDF}}\xspace}  %
\newcommand{\textttbf}[1]{{\texttt{\textbf{#1}}}}
\usepackage{pythonhighlight}
\usepackage{listings}

\definecolor{codegreen}{rgb}{0,0.6,0}
\definecolor{codegray}{rgb}{0.5,0.5,0.5}
\definecolor{codepurple}{rgb}{0.58,0,0.82}
\definecolor{backcolour}{rgb}{0.95,0.95,0.92}

\definecolor{nicegreen}{rgb}{0.,0.5,0.}

\lstdefinestyle{mystyle}{
    backgroundcolor=\color{backcolour},   
    commentstyle=\color{codegreen},
    keywordstyle=\color{magenta},
    numberstyle=\tiny\color{codegray},
    stringstyle=\color{codepurple},
    basicstyle=\ttfamily\footnotesize,
    breakatwhitespace=false,         
    breaklines=true,                 
    captionpos=b,                    
    keepspaces=true,                 
    numbers=left,                    
    numbersep=5pt,                  
    showspaces=false,                
    showstringspaces=false,
    showtabs=false,                  
    tabsize=2,
    xleftmargin=0.5in,
    xrightmargin=0.5in
}

\newcommand{\orcid}[1]{\href{https://orcid.org/#1}{\textsuperscript{\includegraphics[width=8pt]{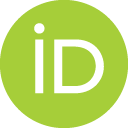}}}}

\begin{document}

\begin{frontmatter}

\title{\Fantomas Unconfined: 
global QCD fits with \Bezier parameterizations}

\author[one]{Lucas Kotz\fnref{fnLK}}
\author[two]{Aurore Courtoy\fnref{fnAC}\corref{cor1}}
\author[three]{T.~J.~Hobbs\fnref{fnTH}}
\author[four]{Pavel Nadolsky\fnref{fnPN}\corref{cor2}}
\author[one]{Fredrick Olness\fnref{fnFO}}
\author[four]{\mbox{Maximiliano Ponce-Chavez\fnref{fnMPC}}}
\author[one]{Varada Purohit}
\affiliation[one]{organization={Department of Physics, Southern Methodist University},%
            city={Dallas},
            postcode={75275-0175}, 
            state={TX},
            country={USA}}
\affiliation[two]{organization={Instituto de Física,
  Universidad Nacional Autónoma de México},%
            addressline={Apartado Postal 20-364}, 
            city={Ciudad de México},
            postcode={01000}, 
            country={Mexico}}
\affiliation[three]{organization={High Energy Physics Division, Argonne National Laboratory},%
            city={Lemont},
            postcode={60439}, 
            state={IL},
            country={USA}}
\affiliation[four]{organization={Department of Physics and Astronomy,
Michigan State University},%
            city={East Lansing},
            postcode={48824}, 
            state={MI},
            country={USA}}
\cortext[cor1]{aurore@fisica.unam.mx}
\cortext[cor2]{nadolsky@smu.edu}
\fntext[fnLK]{\includegraphics[width=10pt]{ORCIDiD_icon128x128.png}~\href{https://orcid.org/0009-0007-5639-0350}{0009-0007-5639-0350}}
\fntext[fnAC]{\includegraphics[width=10pt]{ORCIDiD_icon128x128.png}~\href{https://orcid.org/0000-0002-1825-0097}{0000-0002-1825-0097}}
\fntext[fnTH]{\includegraphics[width=10pt]{ORCIDiD_icon128x128.png}~\href{https://orcid.org/0000-0002-2729-0015}{0000-0002-2729-0015}}
\fntext[fnPN]{\includegraphics[width=10pt]{ORCIDiD_icon128x128.png}~\href{https://orcid.org/0000-0003-3732-0860}{0000-0003-3732-0860}}
\fntext[fnFO]{\includegraphics[width=10pt]{ORCIDiD_icon128x128.png}~\href{https://orcid.org/0000-0001-6799-2436}{0000-0001-6799-2436}}
\fntext[fnMPC]{\includegraphics[width=10pt]{ORCIDiD_icon128x128.png}~\href{https://orcid.org/0009-0003-0139-4072}{0009-0003-0139-4072}}

\begin{abstract}
\Fantomas is a C++ toolkit for exploring the parametrization dependence of parton distribution functions (PDFs) and other correlator functions in quantum chromodynamics (QCD).  \Fantomas facilitates the generation of adaptable polynomial parametrizations for PDFs, called metamorphs, to find best-fit PDF solutions and quantify the epistemic uncertainty associated with the parametrizations during their fitting. The method employs \Bezier curves as universal approximators for a variety of PDF shapes. 
Integrated into the \xFitter framework for the global QCD analysis, \Fantomas provides a foundation for general models of PDFs, while reducing the computational time compared to the approaches utilizing traditional polynomial parametrizations as well as providing an interpretable alternative to neural-network-based models. 
This paper outlines the structure and practical usage of the \Fantomas toolkit, including its inputs, outputs, and implementation within \xFitter. It also provides a practical example of using \Fantomas for uncertainty quantification as well as the combination of PDF fits into a single ensemble.
\end{abstract}

\begin{keyword}
hadron structure \sep parton distributions \sep \Bezier curves \sep universal approximation 
\\
\noindent \textbf{PROGRAM SUMMARY}

\noindent
{\em Program Title: Fantômas}                                          \\
{\em CPC Library link to program files:} (to be added by Technical Editor) \\
{\em Developer's repository link:} \url{https://gitlab.com/cteq-tea/public/fantomas4qcd} \\
\url{https://gitlab.cern.ch/fitters/xfitter/}  in the \textttbf{Fantomas} branch.\\
{\em Licensing provisions(please choose one):} CC0 1.0/CC By 4.0/MIT/Apache-2.0/BSD 3-clause/BSD 2-clause/GPLv3/GPLv2/LGPL/CC BY NC 3.0/MPL-2.0  \\
{\em Programming language: C++}                                   \\
{\em Supplementary material:}                                 \\
{\em Nature of problem:} Dependence on functional forms of parton distribution functions (PDFs) in a hadron introduces a significant uncertainty in precision predictions for many high-energy scattering processes. A new generation of global analyses of PDFs requires streamlined, fast generation of diverse functional forms satisfying number and momentum sum rules consistent with QCD, asymptotic limits, integrability of PDFs, and positivity of the resulting cross sections. Other approaches to approximate a variety of PDFs may lack parsimony and interpretability attributed to explicit functional forms.  \\
{\em Solution method:} The \Fantomas toolkit provides a framework to construct families of increasingly flexible parametrizations for PDFs using \Bezier curves. These polynomial functions are universal approximators that are computed from the values of the function itself at user-specified control points. \Fantomas provides a method to cast many existing PDF parametrizations into an internal format (a metamorph) that simplifies raising the degree of the polynomial solely by adding new control points, without refitting all PDF parameters from scratch.  The \Fantomas approach systematizes exploration of the multidimensional parameter landscape of PDFs while potentially reducing human involvement and computing time compared to traditional realizations. \\
{\em Additional comments including restrictions and unusual features (approx. 50-250 words):}\\

Preprint numbers: ANL-197925~\\

\end{keyword}

\end{frontmatter}

\tableofcontents

\section{Introduction \label{sec:Intro}}
The phenomenology of hadron structure is a bourgeoning subdiscipline of quantum chromodynamics (QCD) linking strong interactions in low- and high-energy domains. Determination of parton distribution functions (PDFs) in nucleons, mesons, and nuclei is advancing rapidly with the influx of  experimental measurements, multiloop theoretical calculations, and lattice QCD computations of quantities which can be related to the PDFs \cite{Amoroso:2022eow,Kovarik:2019xvh,Ethier:2020way}. While the dependence of a PDF $f_a(x,Q)$ on the factorization energy scale $Q$ is precisely specified by perturbative QCD, its dependence on $ x$ --- the momentum fraction  that a parton $a$ carries inside the parent hadron --- is fully non-perturbative and is currently inferred either from experimental data using the method of the global QCD analysis \cite{Morfin:1990ck} or, still limitedly, from a lattice QCD computation. 
The functional forms of PDFs are {\it a priori} unknown locally, while
globally they satisfy several expectations such as sum rules and integrability of moments. 
Search for such functions by fitting them to cross section data has been tackled within two main paradigms: either inferring parameters of user-constructed fixed parametrization forms or using neural networks. 
Proton and other PDFs, so determined, are then used to predict further observables and extract Standard Model parameters. Their role in precision physics is of the utmost importance. It is why the leading groups that determine the most precise proton PDFs \cite{Hou:2019efy,Bailey:2020ooq,NNPDF:2017mvq} joined their efforts in a benchmarking and combination exercise to provide PDFs for the high-luminosity LHC era~\cite{PDF4LHCWorkingGroup:2022cjn}. Further progress towards comprehensive NNLO and N3LO PDFs has been achieved recently by several groups~\cite{NNPDF:2021njg,McGowan:2022nag, NNPDF:2024nan,Ablat:2024muy,MSHT:2024tdn}.

As we strive for ever-increasing precision and accuracy of PDFs, it is essential to account for all leading sources of their uncertainty. Among these, quantifying the important epistemic uncertainty due to the PDF parametrization poses a special challenge in light of the vast number of possible realizations as well as subtle, yet consequential biases that may affect the analysis~\cite{Courtoy:2022ocu}. This viewpoint, and the critical  effort that it implies, has been adopted and refined in further AI/ML-oriented studies~\cite{Kriesten:2023uoi, Kriesten:2024are, Kriesten:2024ist, Kriesten:2025gti}. Assigning a probability distribution to the epistemic uncertainty focuses on the representativeness of the sampling of the solutions for PDFs~\cite{Courtoy:2022ocu,Yan:2024yir,Costantini:2024wby, Costantini:PDF4LHC24}. 

\Fantomas is a C++ package to parametrize PDFs, or $x$-dependent nonperturbative correlator functions in general, with the help of \Bezier curves -- representations for arbitrary polynomials defined in a universal Bernstein basis. \Bezier curves here refer to 1-dimensional polynomials $f(x)$ whose coefficients are computed from the values $f(x_i)$ at several control points $x_i$ specified by the user. 
According to the Stone-Weierstrass theorem -- a classical universal approximation theorem akin to the ones \cite{cybenko_approximation_1989, hornik_universal_1990, hornik_approximation_1991} that justify neural networks -- for any continuous differentiable function of $x$, there is a \Bezier curve that approximates this function to the desired accuracy. 
Such polynomial approximators offer advantages compared to neural-network models, such as low dimensionality of the parameter space and transparent interpretation. 
By constructing the polynomials from their values at control points, instead of fitting the polynomial coefficients directly, one gains a certain leverage over the interpolation or fitting procedure, as explored in this article.

The capabilities of the \Fantomas framework  were first demonstrated in a fit of pion PDFs~\cite{Kotz:2023pbu,Kotz:2025lio}, yielding significant physics insights. The pion PDFs are less known than the nucleon ones. Previous modern analyses of the PDFs in the pion~\cite{Barry:2018ort, Novikov:2020snp,Barry:2021osv} displayed narrow uncertainty bands in spite of the paucity of data in the intermediate $x$ region and large uncertainty at small $x$. We implemented the \Fantomas parametrizations in the \xFitter fitting framework~\cite{xFitterwebsite}, which was used for one such earlier fit \cite{Novikov:2020snp}, and performed a comprehensive analysis of the parametrization dependence by generating about 100 fits with varied \Bezier parametrizations. In these fits, the degree of a polynomial was raised simply by adding a new control point to the parametrization available from a previous fit. These and other features streamlined generation of distinct fits, which were then combined into a single uncertainty band using the METAPDF combination method \cite{Gao:2013bia}. The uncertainty from the nuclear PDFs was added in the final release of \FantoPDF PDFs \cite{Kotz:2025lio}. This was the first set of pion distribution functions to comprehensively account for the parametrization uncertainty, achieving notable agreement with lattice QCD determinations of the pion PDFs. A key finding was that the large-$x$ experimental data constrains the valence PDF, whose fixed-order perturbative behavior is consistent across fitting groups. In contrast, the sea and gluon PDFs are anti-correlated and remain challenging to disentangle due to the limited data available at small $x$.  
\\

The \Fantomas toolkit is shared on our git repository in two versions. First, a standalone version demonstrates core features of the \Fantomas methodology on simple examples and does not have any significant external dependencies. The \xFitter version is built as an extension of the public \xFitter code. Specifically, we exploited the modular structure of \xFitter 2.2.0 to integrate the \Fantomas components by adding a section in the module with PDF parametrizations. As a validation example, in the \xFitter version we included a short fit of the pion PDFs from our baseline analysis in \cite{Kotz:2023pbu}. Both versions will be described hereafter.
In this paper, as in Ref.~\cite{Kotz:2023pbu}, we have illustrated the functionality of \Fantomas using a toy model in 
\textsf{Mathematica} (available by request) and the \xFitter set-up for the pion PDF fit~\cite{Novikov:2020snp}, hence addressing both simple and real physics-case scenarios.

The paper is structured as follows. The core of the problem we confront through \Fantomas, namely the role of parametrizations in global analyses, is motivated in Section~\ref{sec:func_form}. The  mathematical concept behind the \meta solution to this problem, as well as first-principle constraints on PDFs, are spelled out in that same Section.
The \Fantomas toolkit is fully detailed in Section~\ref{sec:fantomas}, where we discuss the main code base: classes, core source files, and steering cards, along with several examples.
This section also includes the inputs and outputs, as well as the various options to activate the \meta, such as the nature of the control points. Then, both the standalone version and the \xFitter implementation are discussed.
We then illustrate in Section~\ref{sec:FantoPDFs} how to use \Fantomas in the context of pion PDF fits. This includes a step-by-step procedure (Section~\ref{sec:0thfit}) for generating the fits; a dissection (Section~\ref{sec:PionFits}) of the resulting ensemble of analyses intended to 
illustrate the behavior of \Fantomas; and an exploration (Section~\ref{sec:flexibility}) of the full power of \meta through guidelines on tuning the flexibility of the parametrization(s).
Knowing  the full power of the \meta, we discuss the combination of PDFs from various parametrizations as well as with further, uncorrelated, uncertainties in Section~\ref{sec:combination}. 
After summarizing the ``do's and dont's" in Section~\ref{sec:guidelines}, we draw our conclusions in Section~\ref{sec:conclusions}.
The various Appendices offer more details on technical aspects of the code and its performance, from a brief review of the \xFitter environment (\ref{sec:xfit}),
to a discussion (\ref{sec:num}) on numerical artifacts pertaining to the pion fits of Section~\ref{sec:FantoPDFs},
as well as a sample \texttt{Mathematica} notebook (\ref{sec:mathematica}). 
\section{Functional forms for PDFs}
\label{sec:func_form}

\subsection{The role of parametrization studies in modern PDF fits \label{sec:RoleOfParametrizationStudies}}

In various inverse problems arising in global analyses of hadron structure, the inference of unknown functions quantifying
non-perturbative QCD dynamics starts by choosing a convenient (non)parametric representation, which must
be differentiable and satisfy certain physics limits. This representation is then constrained by a fit to discrete data. 
Among the various representations, polynomial functions and neural networks (NNs) are most commonly used and have reached
the highest level of sophistication in QCD analyses of unpolarized PDFs.
In practice, modern PDF fits must continually revisit the question of the parametric flexibility which is needed in order to adequately model an ever-expanding global set of hadronic data. Global fits based on explicit, analytical parametrizations typically operate with polynomial-based functional forms with 5-10 parameters per flavor, mirroring the adaptive nature of [trained and cross-validated] NNs. Meanwhile, NN-based fits must similarly confront their inherent parametrization dependence as realized through their chosen number of nodes, hidden layers, and overall network topology.
Addressing these challenges in a systematic fashion requires the nontrivial exploration of a wide array of parametrization forms and hyperparameters.
The need for quantifying this parametrization dependence derives from the fact that the effect of choosing an assumed functional forms for analytical parametrizations or, equivalently, the architecture and training of a NN, can be substantial in precision PDF fits to large data sets, contributing together with other sources \cite[Sec.~1G in][]{Kovarik:2019xvh} to the total PDF uncertainty. %

Within this context, the \Fantomas methodology we present in this manuscript streamlines the generation and fitting of versatile polynomial functional forms, thereby further narrowing the performance gap between these two standard approaches to parametrization in PDF analyses while supplying a novel approach for the fast generation of many explicit fitting forms.
Polynomial parametrizations are appealing in that, while they are able to approximate arbitrary well-behaving functions, they have low complexity %
and are %
amenable to algebraic derivation from various predictions using methods of classical statistics, approximation theory, or the nonperturbative modeling of hadron structure.
When we fit such a parametrization, we generally expect a tradeoff between the low agreement with the data, when the parametrization is too inflexible, and loss of the generalization power when the parametrization is too flexible and responsive to the stochastic noise in the data. Training of neural networks often employs cross validation to obtain a PDF model that reasonably balances between these two extremes. The point of the balance can be approximately determined, {\it e.g.}, by controlling the bias-to-variance ratio \cite{NNPDF:2021njg} in either the analytical or NN approach. 

That said, a large volume of data does not imply a unique PDF model or full suppression of the parametrization uncertainty.  
As a matter of fact, functional forms for PDFs cannot be uniquely determined solely based on discrete experimental measurements because of the mathematical property of PDF parameterization mimicry~\cite{Courtoy:2020fex}, which is enhanced in the regions of scarce data coverage, known as extrapolation regions. The non-uniqueness of the best-fit solution can be demonstrated first by interpolating or fitting data with a polynomial approximator and applying the unisolvence theorem \cite{Courtoy:2020fex}, and then showing that the result extends to other types of universal approximators, which themselves can be approximated by some polynomial to a given accuracy. 

A closure test of a given fitting methodology \cite{NNPDF:2021njg, Harland-Lang:2024kvt}, verifying an excellent agreement between the best-fit PDF solution and the known ``truth PDF" used to generate the fitted pseudodata, does not generally preclude existence of multiple functional forms achieving comparable agreement with the discrete data when such ``truth" is {\it a priori} unknown. In the Bayesian paradigm, one must initially assume an uninformed prior when trying equally plausible PDF functional forms. The combined posterior prediction based on the totality of explored functional forms can then be constructed using several available methods, such as Bayesian model averaging \cite{Jay:2020jkz,Yan:2024yir} or discrete profiling \cite{Dauncey:2014xga}. All such studies run into a serious challenge of representative and unbiased exploration (sampling) of large parametric spaces for PDFs \cite{Courtoy:2022ocu} in order to minimize the distributional (miscoverage) component of the epistemic uncertainty \cite{Malinin:2018, Kriesten:2023uoi, Kriesten:2024are, Kriesten:2024ist}.
Such exploration is systematized through the use of \Bezier curves to generate various parametrizations on the fly~\cite{Kotz:2023pbu}.

\subsection{Metamorph}
\label{sec:metamorph}

\begin{wrapfigure}{BTH}{0.5\textwidth}
  \begin{center}
    \includegraphics[width=0.49\textwidth]{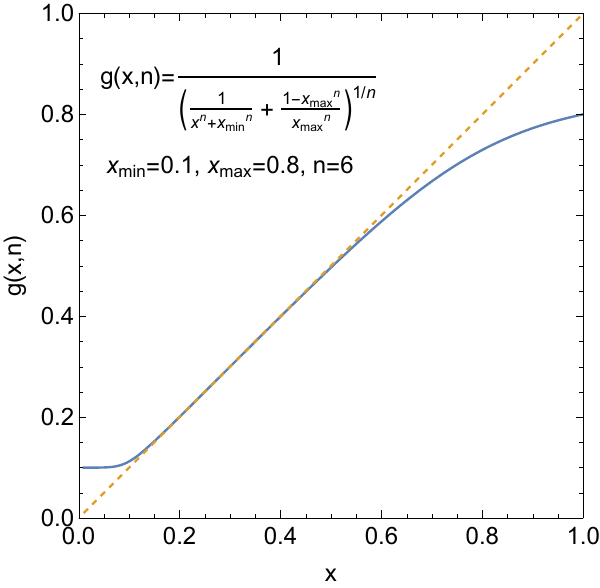}
  \end{center}
  \caption{Plot of the stretching function $g(x,n)$ in Eq.~(\ref{eq:stretch}) for $x_\textrm{min}=0.1$, $x_\textrm{max}=0.8$, $n=6$. The dashed curve indicates the dependence $y=x$.  }
  \label{fig:scaling}
  \vspace{0.5cm}
\end{wrapfigure}

The methodology of \Fantomas is based on the use of analytical forms called {\it metamorphs} to represent functions of a variable $x$ defined over a supported region, $x\in [0,1]$. In the context of PDFs, such functions represent the distribution, $f_a(x,Q^2_0)$, of a parent hadron's fractional momentum, $x$, carried by a constituent parton of flavor, $a$, at an starting scale, $Q^2_0$, of the 
Dokshitzer-Gribov-Lipatov-Altarelli-Parisi (DGLAP) evolution. This Section describes the mathematics behind the \meta approach, while subsequent illustrations will be given throughout the article.

PDFs must follow, at small and large $x$, physically-motivated constraints and analyticity.
To accommodate for such end-point behaviors,  we parametrize  the PDF as
\begin{equation}
    xf_a(x,Q_0^2)=F_a^{car}(x)\times F_a^{mod}(x)
    \label{eq:metamorphbasis}
\end{equation}
\noindent for each parton flavor $a$ contributing at scale $Q_0$. The ``baseline" of our functional form is called the \textit{carrier function}, $F_a^{car}$. The carrier function serves as the simplest parameterization form consistent with the above-mentioned constraints, particularly under the scenario $F_a^{mod}(x)\! =\! 1$. We have chosen to define our carrier function as
\begin{equation}
    F_a^{car}(x)\equiv A_ax^{B_a}(1-x)^{C_a}.
    \label{eq:carrierbasis}
\end{equation}
\noindent Some normalization coefficients, $A_a$, are chosen to satisfy the flavor number and momentum sum rules; the other  remain as free parameters of the fit.
$B_a$ and $C_a$ govern the behavior of the parametrization under the limits $x\rightarrow 0$ and 1.
Conceptually, the purpose of the carrier is to implement expectations for the $x$ behavior following from the first principles or semi-qualitative models.

 In Eq. (\ref{eq:metamorphbasis}), $F_a^{mod}$ is the \textit{modulator function}. The modulator allows more freedom at $x$ away from 0 or~1. The modulator, expressed as $F_a^{mod}\left(x;{\cal B}^{(N_m)}\right)$, is defined to be a function of ${\cal B}^{(N_m)}(y)$, a polynomial of order $N_m$ and argument $y\equiv y(x)$, where $y(x)$ is a \textit{stretching function}. In the simplest realization, $y(x)=x$. However, it can be any function of $x$. As an example, in the pion study \cite{Kotz:2023pbu}, the modulator and stretching function are defined as
\begin{equation}
    F_a^{mod}\left(x; {\cal B}^{(N_m)}  \right)
   =  
 1+ {\cal B}^{(N_m)}(y),
     \label{eq:modulator}
\end{equation}
\noindent and
\begin{equation}
    y=(g(x,n))^{\alpha_x}\equiv \left( \frac{1}{\left( \frac{1}{x^n+x_{\rm min}^n} + \frac{1-x_{\rm max}^n}{x_{\rm max}^n}\right)^{1/n}} \right)^{\alpha_x},
\label{eq:stretch}
\end{equation}
\noindent with $0 < \alpha_x < 1$ and $n=6$~\cite{Kotz:2023pbu}, as illustrated in Fig.~\ref{fig:scaling}. The function $g(x,n)$ freezes the $x$ dependence below $x_\textrm{min}$ and above $x_\textrm{max}$, where both $x_\texttt{min}$ and $x_\texttt{max}$ lie in the extrapolation regions of $x$ (outside of the range covered by the data). In other words, 
\begin{equation}
    \lim_{x\ll x_\textrm{min}} g(x,n) = x_\textrm{min}, \quad \lim_{x\gg x_\textrm{max}} g(x,n) = x_\textrm{max}, 
\end{equation}
{\it i.e.}, the asymptotic behavior of the whole \meta $xf_a(x,Q_0^2)$ is controlled by the carrier $F_a^{car}(x)$ in these asymptotic limits. \Fantomas automatically sets $x_\textrm{min}$, $x_\textrm{max}$ according to the mandatory fixed control points, as described in Sec.~\ref{sec:0thfit}. The power $n$ determines the steepness of the freezout transition at $x\sim x_\textrm{min}$ and $x\sim x_\textrm{max}$. In between $x_\textrm{min}$ and $x_\textrm{max}$, $g(x,n) \approx x$, cf. Fig.~\ref{fig:scaling}, so that $y\approx x^{\alpha_x}$. Hence, in the $x$ region covered by the data, the argument of the \Bezier curve is essentially $x^{\alpha_x}$, where commonly used parametrizations take $\alpha_x = 0.5$, 1, or another positive fractional value.  

In our methodology, the $\cal B$ polynomial is chosen to be a \Bezier curve, 
\begin{equation}
{\cal B}^{(N_m)}(y) \equiv  \sum_{k=0}^{N_m} c_k\ B_{N_m,k} (y)
\label{eq:BezierCurve}
\end{equation}
by introducing a basis of Bernstein polynomials,
\begin{equation}
    B_{N_m,k} (y)\equiv \left(\begin{matrix} N_m  \\ k  \end{matrix}\right) y^k (1-y)^{N_m-k} \ ,
    \label{eq:BNml}
\end{equation}
where the $\binom{N_m}{k}$ is a binomial coefficient. The Bernstein polynomials are neither orthogonal nor oscillatory. For a fixed degree $N_m$ and index $k \in \{0, 1, \ldots, N_m\}$, the Bernstein basis polynomial $B_{N_m, k}(y)$ attains its unique global maximum at $y_{(k)} = k/N_m$. Thus, the choice of the stretching function $y(x)$ controls spacing of the peaks $y_{(k)}$ of the Bernstein polynomials over $x\in [0,1]$.

With the \Bezier curve, the numerical coefficients $c_k$ in Eq.~(\ref{eq:BezierCurve}) are not fitted directly; 
instead, they are calculated from the values of the modulator function $F_{a}^{mod}\left(x_i; {\cal B}^{(N_m)}  \right)$ at $N_{m}+1$ distinct points $x_i$ in the interval $x\in [0,1]$, called \textit{control points} (CPs). In the C++ code, class \texttt{ControlPoint} refers to a pair of real numbers $\{x, {\cal B}^{(N_m)}(y(x))\}$ specifying the $x$ position (fixed during the fit) and the corresponding \Bezier value, which can be a free parameter, for every CP.
Hence, in contrast to traditional parametrizations used by PDF fits, the free parameters controlling the modulator are the values of the \Bezier curve, {\it i.e.}, effectively the modulator $F_a^{mod}=1+{\cal B}^{(N_m)}$, at user-specified CPs. 

Denote $P_i\equiv {\cal B}^{(N_m)}(y(x_i))$ at a control point $x_i$.
We will solve for $c_k$ in terms of $P_i$ a matrix form \cite{Farin:2001,pomax}, denoting a scalar, an $(N_m+1)$-dimensional vector, and an $(N_m+1)\times (N_m+1)$ matrix as $S$ (no underline), $\underline V$, and $\underline{\underline M}$,  respectively.  We use Einstein's summation with Roman indices taking values between 0 and $N_m$. The unisolvence theorem for polynomials guarantees existence of a unique solution for $c_k$ under the assumption that the Vandermonde matrix $T$ defined below is invertible. 

Given the vector ${1,\ y,\ y^2,\ldots,\ y^{N_m}}$ containing incrementing integer powers $y^p$ as $\underline Y (y)$, the vector of ${\cal B}^{(N_m)}(y_i)$ at $(N_m+1)$ CPs $y_i \equiv y(x_i)$ as $\underline P$, and the vector of the Bernstein polynomial coefficients $c_k$ as $\underline C$, we can show that our \Bezier curve ${\cal B}^{(N_m)}(y)\equiv {\cal B}$ in Eq.~(\ref{eq:BezierCurve}) assumes the form
\begin{equation}
  {\cal  B} =  {\underline Y}^T(y) \cdot \underline{\underline M} \cdot {\underline C}, 
  \label{eq:Bmatrix}
\end{equation}
where $\underline{\underline M} $ is an $(\! N_m \!
+ \! 1 \!) \times (\! N_m \! + \! 1 \!)$ matrix 
 containing combinations of binomial coefficients, 
 \begin{equation}
\underline{\underline M}\equiv \|m_{lp}\| \mbox{ with } m_{lp}=\begin{cases}
(-1)^{p-l}\left(
\begin{array}{c}
l\\
n
\end{array}\right)
\left(\begin{array}{c}
n-p\\
n-l
\end{array}\right), &  l\leq p\\
0, & l>p
\end{cases}.
\label{M}
\end{equation}

From Eq.~\ref{eq:Bmatrix}, we can represent $\underline P$ as 
\begin{equation}
     {\underline P}={\underline{\underline T}}\, \cdot {\underline{\underline M}}\, \cdot {\underline C}
     \label{eq:Pmatrix}
\end{equation}
by introducing the Vandermonde matrix  $T\equiv \|Y_k(y_i)\| $ with $Y_k(y_i)=(y_{i})^{k}$ ($y_i$ raised to power $k$). 
Since under our assumptions it is a non-singular square matrix, we can invert this equation to obtain vector $\underline C$ of \Bezier coefficients as a function of CPs, 
\begin{equation}
  \underline C = \underline{\underline M}^{-1} \cdot \underline{\underline T}^{-1} \cdot {\underline P}.
  \label{eq:Cmatrix}
\end{equation}
Once $\underline C$ is found, we reconstruct the modulator $F_a^{mod}\left(x; {\cal B}^{(N_m)}  \right)$ at any $x \in [0,1]$ using Eqs.~(\ref{eq:modulator}) and (\ref{eq:Bmatrix}), and then the complete \meta $xf_a(x,Q_0^2)$ in Eq.~(\ref{eq:metamorphbasis}). 

The \Bezier curve calls for the inversion of matrix $\underline{\underline T}$ in Eq.~(\ref{eq:Cmatrix}), which requires some care, especially at high $N_m$. \ref{sec:num} discusses the optimal placement of CPs that leads to stable inversion.

Many polynomial parametrizations used in PDF global analyses can be cast into the \meta form simply by recomputing their numerical coefficients (without refitting). It is the case for the parametrizations utilizing the Bernstein basis used by CT, as well as the Chebyshev basis used by MSHT. While the neural networks of NNPDF are employed in a conceptually different manner, they could still be linked to polynomial approximators, given the smoothness of respective PDF solutions. 

\subsection{Sum rules and prior constraints}
\label{sec:sum_rules}

The form of a metamorph is usually constrained by global considerations such as sum rules and prior probabilities. In the current \Fantomas toolkit, we implemented support for two main sum rules imposed on the unpolarized PDFs -- the valence and the momentum sum rules, related, respectively, to the $0^{th}$ and the $1^{st}$ Mellin moment,
\begin{equation}
    \langle x^n f_a \rangle (Q^2) = \int_0^1 \, x^n\, f_a(x, Q^2),
    \label{eq:mellin_mmt}
\end{equation}
with $q$ representing the flavors or flavor combinations, {\it e.g.}, $q=\{V, \, S, \, g\}$ for a pion target \cite{Novikov:2020snp,Kotz:2023pbu}.
To uphold the sum rules in a given decomposition, the first moments need to be obtained through numerical integration. These sum rules usually dictate the value of the normalization for one \meta function each. For example, in the pion study~\cite{Kotz:2023pbu}, the two sum rules to be fulfilled are
\begin{equation}
    \langle V \rangle = 2,
    \label{eq:valencesumrule}
\end{equation}
\noindent and the momentum sum rule,
\begin{equation}
    \langle xV \rangle +  \langle xS \rangle +  \langle xg \rangle = 1.
    \label{eq:momentumsumrule}
\end{equation}

The \texttt{metamorph} class \Fantomas provides a method \texttt{GetMellinMoment} to compute arbitrary Mellin moments at scale $Q_0$ by numerical integration. As a byproduct, the method tests for integrability of the \meta function given its numerical parameters. 
Other prior constraints are related to analyticity of PDFs, smoothness and positivity (of cross sections). The latter can be tracked through prior penalties on the value ranges of specific parameters, as will be made explicit in Section~\ref{sec:penal}.

\section{The \Fantomas toolkit}
\label{sec:fantomas}

The \Fantomas toolkit described in this Section provides a \texttt{C++} implementation of \meta parametrizations for PDFs detailed in Sec.~\ref{sec:func_form}. It also includes a wrapper in \texttt{C} that facilitates compilation with \texttt{Fortran} programs such as \xFitter.

We review the classes of the \Fantomas toolkit in Sec.~\ref{sec:Classes}; core files, inputs, and outputs in Sec.~\ref{sec:Fantofiles}, including an example of the steering card in Sec.~\ref{sec:steering_card}; the fitting algorithm in Sec.~\ref{sec:algorithm}; the C wrapper functions in Sec.~\ref{sec:CWrapper}; the standalone implementation in Sec.~\ref{sec:standalone}; and the \xFitter implementation in Sec.~\ref{sec:xFitterImplementation}.

\subsection{Classes \label{sec:Classes}} 
\Fantomas defines the following classes:

\begin{enumerate}
    \item Class \texttt{ControlPoint} defines a control point at position \texttt{x}, with the value of the function at this point accessed via a pointer \texttt{*ps}. 
     \item Class \texttt{metamorph} implements a PDF parametrization for a single flavor according to Eq.~(\ref{eq:metamorphbasis}) in the form
     \begin{displaymath}
\texttt{xf(x) = Carrier(x, Sc)*Modulator(x, Sm)},
  \end{displaymath}
 where \texttt{Carrier} defines the asymptotic behavior in the limits $x\to 0$ and $x\to 1$; \texttt{Modulator} determines the behavior of f(x) over the interval $0 < x < 1$. Parameters of the \texttt{Carrier} and \texttt{Modulator} are respectively stored in external arrays \texttt{Sc}, containing $A_a$, $B_a$, $C_a$ in Eq.~(\ref{eq:carrierbasis}), and \texttt{Sm} with values of $P_i$. Internally, \texttt{Modulator(x, Sm)} is controlled by a private vector of \Bezier coefficients $\underline C$. Every time when the fit updates the CP values $P_i$, {\it i.e.}, $F^{mod}(x_i)-1$, \Fantomas must recompute $\underline{C}$ by calling \texttt{metamorph::UpdateModulator()}.  \vspace{1\baselineskip}
 
 The \texttt{\meta} class provides the following public methods:
 \begin{itemize}
    \item \texttt{double metamorph::f(const double x)} returns the value of the \meta for the momentum fraction $x$. 
    \item \texttt{void metamorph::UpdateModulator()} computes vector $\underline C$ of \Bezier coefficients using Eq.~(\ref{eq:Cmatrix}). This computation must be done every time after the control points ({\it i.e.}, the array \texttt{Sm[]}) are updated, and before the metamorph value \texttt{f} is computed. 
     \item \texttt{double metamorph::Modulator(const double x)} returns the value of the Modulator function for the momentum fraction $x$.
     \item \texttt{double metamorph::yx(const double x)} returns the stretched argument $y(x)$ of the \Bezier curve.
      \item \texttt{void metamorph::SetXstretching()} and \texttt{void metamorph::GetXstretching()} set and return parameters of the stretching function $y(x)$.
     \item \texttt{double metamorph::GetMellinMoment(double MellinPower, int npts = 10000)} returns $\langle x^{n+1} f_a\rangle$, with $n=\texttt{MellinPower}$, {\it i.e.}, the $(n+1)$-th Mellin moment over $0 \leq x \leq 1$. This prescription for orders of Mellin moments follows the one adopted in xFitter. \texttt{npts} is the number of points for Gaussian integration over $x$. By default, \texttt{npts=10,000}.
      \item \texttt{double metamorph::Cs(const int i)} returns the value for the \texttt{i}-th \Bezier coefficient.
     \item \texttt{double metamorph::GetConditionNumber()} returns the condition number of matrices $T$ and $T^{-1}$ using the Frobenius norm, $||T||$, {\it i.e.}, the square root of the sum of absolute squares of the elements. The condition number is calculated as $||T|| \cdot ||T^{-1}||$. See \ref{sec:num}. 
      \item \texttt{double metamorph::chi2prior()} returns a contribution to $\chi^2$ from the prior imposed on the metamorph

 \end{itemize}

\item Class \texttt{metamorphCollection} is a container for \texttt{Nmeta} metamorph objects providing parametrizations for parton distributions of distinct flavors. \texttt{metamorphCollection} also provides methods to read, update, and return parameters and PDF values for a metamorph object corresponding to a physical flavor ID \texttt{ifl}, such as the PDF flavor assignment according to the LHAPDF convention. The physical flavor ID is specified in the steering card, cf. Sec.~\ref{sec:steering_card}, and matches the flavor assignment in the external fitting program. \vspace{1\baselineskip}

Private members of \texttt{metamorphCollection} include: 
\begin{itemize}
    \item \texttt{vector<metamorph> MetaVector} contains metamorph objects in the \texttt{metamorphCollection};
    \item \texttt{map<int, metamorph*> MetaRoster} maps the integer physical flavor \texttt{ifl} onto the pointer to the corresponding metamorph in \texttt{MetaVector};
  \item \texttt{map<int, int> PositionRoster} maps the physical flavor \texttt{ifl} onto the integer position \texttt{iMet} of the corresponding metamorph in \texttt{MetaVector}. 
\end{itemize}\vspace{1\baselineskip}

 The \texttt{metamorphCollection} class provides the following public methods:
\begin{itemize}
    \item \texttt{void metamorphCollection::ReadCard()} reads input parameters (powers of the carrier, control points, etc.) for all metamorphs from an input steering card, see Sec.~\ref{sec:steering_card}. Analogously, \texttt{metamorph Collection::WriteCard()} writes all parameters into a new steering card.
  \item \texttt{void metamorphCollection::UpdateParameters(const int ifl, double *deltas} updates the metamorph for physical flavor \texttt{ifl} according to the provided changes \texttt{deltas} in the parameters. \texttt{UpdateModulator()} is called within this method.
  \item \texttt{double metamorphCollection::f(const int ifl, const double x)} returns the value of a metamorph with physical flavor \texttt{ifl} at momentum fraction \texttt{x}. \\
  \texttt{metamorphCollection::MellinMoment} and \texttt{metamorphCollection::GetCondition\allowbreak Number} respectively return a Mellin moment and condition number for the metamorph with physical flavor \texttt{ifl}. 
  \item \texttt{int metamorphCollection::GetMetamorphCount()} returns the number of metamorph members in \texttt{MetaVector}.
  \item \texttt{double metamorphCollection::chi2prior()} returns a contribution to the objective function, $\chi^2$, from the sum of priors on all metamorphs
  \item \texttt{\small void metamorphCollection::MetamorphDiagnostics(const string\& outputDirectory)} saves diagnostics about the metamorph collection into \texttt{fantomas\_functional\_parameters.txt} in a directory \texttt{outputDirectory}. The diagnostics includes values of carrier and modulator parameters, momentum fractions, and condition numbers. \texttt{void metamorphCollection::Chi2Diagnostics} saves the diagnostics about prior penalties. In \xFitter, \texttt{outputDirectory = output/}. 
   \end{itemize}    
\end{enumerate}

\subsection{The core source files, steering cards}
\label{sec:Fantofiles}

The following \textbf{source files} and the respective C++ header (\texttt{.h}) files define the \Fantomas classes and provide core functionalities:
\begin{enumerate}
        
    \item \texttt{metamorph.cc} defines classes \texttt{ControlPoint} and \texttt{metamorph} for individual control points and \meta parametrizations, respectively.
    \item \texttt{metamorphCollection.cc} defines class \texttt{metamorphCollection}. A companion file\\ \texttt{metamorphCollectionPrior.cc} contains definitions of the prior penalties on metamorphs that can be fit-specific and saved in \xFitter together with the rest of the outputs.
     \item \texttt{LUPinverse.cc} and \texttt{cl2DArray.h} support matrix operations, including inversion.
     \item \texttt{FantoWrapper.cc} provides an \texttt{extern "C"} wrapper for the core \Fantomas methods,  see Sec.~\ref{sec:CWrapper}. 
\end{enumerate}

\textbf{Steering cards} play an important role in the \Fantomas algorithm, in that they store metamorph parameters and settings in an extended format that is not normally supported by control cards in \textsf{MINUIT} or another fitting program. During the fitting, the output steering card saves the metamorph parameters that achieved the lowest $\chi^2$ among all values explored. We set the \textsf{MINUIT} fitting parameters to correspond to the changes from the tabulated values in the steering card, see Sec.~\ref{sec:algorithm}, which simplifies bookkeeping. \Fantomas therefore operates with two steering cards:

\begin{itemize}
    \item \texttt{steering\_fantomas.txt}: an input file that contains the initial parameters -- information about the control points, order of the \Bezier curves, the mapping parameters-- used to create the \meta. This file is accessed by the \texttt{void metamorphCollection::ReadCard()} method.
    \item \texttt{steering\_fantomas\_out.txt}: this is a new steering card containing the updated \meta parameters. It is created by calling \texttt{metamorphCollection::WriteCard()}. In the \xFitter realization, this card is updated every time when a lower $\chi^2$ is found. 
\end{itemize}
\subsubsection{Example of a steering card}
\label{sec:steering_card}

The steering card stores parameters and settings for all \meta functions. Figure~\ref{fig:steeringcard} shows an example of such a card. Line 1 specifies the format version and should not be modified. The header with comments marked by ``\#" is followed by several blocks for input metamorphs,  with each metamorph associated with its physical flavor ID \texttt{ifl} (according to the LHAPDF or another convention), 
by which this metamorph can be addressed from the external code via the respective method of \texttt{metamorphCollection}.
 For each such block, the settings are 
\begin{displaymath}
\{N_m,\ \mathtt{ MappingMode},\ \mathtt{ xPower}\equiv \alpha_x,\ \mathtt{Sc[0]}\equiv A_a,\ \mathtt{Sc[1]}\equiv B_a,\ \mathtt{Sc[2]}\equiv C_a\}.    
\end{displaymath}
The first entry is the degree of the polynomial; the second is related to restricting the value of $f(x)$ to stay within constraints -- the mapping mode. It is a placeholder for imposing bounds on the parametrization.  
 Currently we only use $\texttt{MappingMode}=0$, which does not impose constraints. The scaling factor $\mathtt{xPower}$ is defined in Eq.~(\ref{eq:stretch}), and the array $\texttt{Sc[]}$ contains the carrier's parameters in Eq.~(\ref{eq:carrierbasis}).
Then, the block contains $(N_m+1)$ records with parameters for the metamorph's CPs, $\mathtt{Xs}\equiv x_i$ for the $x$ positions, and $\mathtt{Sm}[i]$ for either $P_i$ or the flag indicating a CP of a special type. 

Specifically, the $\mathtt{Sm}$ column should contain one of the following four entries to select from:
\begin{itemize}
    \item A real number (referred to as ``FREE" in~\cite{Kotz:2023pbu}): this is the initial $P_i={\cal B}^{(N_m)}(y_i)$ value of the \Bezier function used in the \meta.\footnote{We remind the reader that the full \meta at a CP $x_i$ is given by $F^{car}(x_i)\times (1+P_i)$.}
    \item ``FIX": This option fixes $P_i=0$, $F_a^{mod}(x_i)=1$ for the rest of the fit, so that the metamorph {\it always} coincides with the carrier function at this CP. \textbf{We recommend to FIX the zeroth and last CPs, with $x_0$ and $x_{N_m}$}, in which case \Fantomas sets $x_\textrm{min}=x_0$, $x_\textrm{max}=x_{N_m}$ in the stretching function $g(x,n)$ in Eq.~(\ref{eq:stretch}). With this setting, $F^{mod}=1$ at $x<x_0$ and $x>x_{N_m}$: the metamorph then coincides with the carrier $A_a x^{B_a} (1-x)^{C_a}$ at $x\to 0$ or 1.    
    \item ``NEW": This option will calculate $P_i$ at this CP, $x_i$, by using a \meta function of a lower order, {\it e.g.}, $N_m-1$, specified solely by the rest of the provided control points. \\
    \textbf{Use the NEW option to add control points to the pre-existing modulator, hence raising the polynomial degree $N_m$ without modifying the modulator's initial shape or values at the already existing CPs.}
    \item ``CALC": This option will calculate the $P_i$ value at this CP when writing the output. This control point will not be used to calculate the \meta function. 
\end{itemize}
\begin{figure}[h]
\begin{python}
# Fantomas steering card v. 1.0cd 
# YYYY-MM-DD HH:MM:SS
# gluon -- metamorph 0
# ifl  Nm  MappingMode  xPower  Sc(0)  Sc(1)  Sc(2)
0 0    0   1.	1.	0.	0.
# Xs  Sm  
0.7	FIX
# sea  -- metamorph 1
# ifl  Nm  MappingMode  xPower  Sc(0)  Sc(1)  Sc(2)
1 3    0   1.	1.	1.02	9.14
# Xs  Sm  
0.001  FIX
0.4    0
0.7    0
0.8    FIX
# valence  -- metamorph 2
# ifl  Nm  MappingMode  xPower  Sc(0)  Sc(1)  Sc(2)
2 3    0   1.	1.	0.73	0.95
# Xs  Sm  
0.01  FIX
0.4   0
0.7    0
0.999  FIX
\end{python}

\caption{An example of a \Fantomas steering card, \texttt{steering\_fantomas.txt}.\label{fig:steeringcard}}

\end{figure}

\subsection{The fitting algorithm \label{sec:algorithm}}

The fitting in \xFitter proceeds as follows. We create an input steering card, \texttt{steering\_fantomas.txt}, containing the initial metamorph parameters. We also modify a \textsf{MINUIT} steering card to assign free parameters to the {\it changes} of the varied metamorph parameters. 
We call the latter the $\delta$ (delta-) parameters, such that $\{A_a, B_a, C_a, c_{l, i}\}\to \{A_a, B_a^0+\delta B_a, C_a^0+\delta C_a, c_{l, i}^0+\delta c_{l, i}\}$, for all but the $A$'s. Those $A_a$ can either be determined by sum rules or let as free parameters. For example, out of the three normalizations of the pion PDFs, the valence normalization constant $A_V$ is calculated following the valence sum rule,
\begin{eqnarray}
 A_V \int_0^1 f^V_{\rm meta}(x)dx=\langle V \rangle %
    \label{eq:valencenorm}
\end{eqnarray}
where  $f^V_{\rm meta}$ is the output of the \meta module.\footnote{ Another way of enforcing the correct normalization relies on the analytical expression for the first moment. This has been used extensively in the past, and still adopted by the
the \xFitter team in the pion study~\cite{Novikov:2020snp}. However, we chose to not forgo a constant in front of $A_V$ to keep our \meta function as general as possible, and to keep the parametrization linear in the parameters. Comparisons of both methods for the pion PDFs have been discussed in Ref.~\cite{Kotz:2023pbu}.} Either the sea normalization or the gluon normalization is varied independently,  while the third normalization is computed to fulfill the momentum sum rule by subtraction, {\it e.g.}, at $Q_0^2$, 
\begin{eqnarray}
    \langle xg \rangle [A_g(A_V, A_S)] =1-\langle xV \rangle [A_V] - \langle xS \rangle [A_S],
    \label{eq:mmtSR}
\end{eqnarray}
where we have made the dependence on the normalization explicit.

Hence, in the external fitting program, a delta parameter must be assigned to every free parameter of the metamorphs: \texttt{Sc[]} as well as \texttt{Sm[]} for {\it free} CPs. While the deltas for FIXed CP's are redundant and will be ignored by \texttt{metamorph\allowbreak Collection} even if assigned, the structure of the \xFitter input nevertheless requires to assign the delta to the FIXed CP's as well. In the standalone \texttt{MINUIT} or other fitting programs, the deltas can be skipped for the FIXed CPs.

Within \texttt{parameters.yaml} of \xFitter, all delta parameters are initialized with a zero value. They are updated as the fit progresses and passed into \texttt{metamorphCollection} to compute the current \texttt{Sm} and \texttt{Sc} parameters from the sum of the corresponding initial value and the delta.
The final metamorph parameters corresponding to the lowest $\chi^2$ are saved in \texttt{steering\_fantomas\_out.txt}. The final values of deltas are saved only in the diagnostic files, such as \texttt{minuit.out.txt}.

For example, for the change 
\texttt{delBv} in $B_V$ in \xFitter (parameter 7 in \textsf{MINUIT}), we assign in \texttt{parameters.yaml}: 
\begin{python}
Parameters:
...
delBv: [ 0., 0.03]
...
Parameterisations:
  v:
    class: Fantomas
    parameters: [Av,delBv,delCv,delDv,delEv,vifl]
...    
\end{python}
The respective input printout in \texttt{MINUIT.out.txt} is %
\begin{python}
 PARAMETER DEFINITIONS:
    NO.   NAME         VALUE      STEP SIZE        LIMITS
    ...
     7 'delBv     '    0.0000      0.30000E-01     no limits
    ...
\end{python}
The \texttt{MINUIT} output printout, which is only retained for diagnostics, is
\begin{python}
...
   EXT PARAMETER                                   STEP         FIRST
  NO.   NAME        VALUE          ERROR          SIZE      DERIVATIVE
   ...
   7    delBv     -0.30241       0.94832E-01   0.24034E-04  -0.14744E-03
\end{python}
The updated value of $B_V$ is saved as \texttt{Sc[1]} for the valence metamorph in \texttt{steering\_fantomas\_out.txt}.

We can reuse \texttt{steering\_fantomas\_out.txt} as a steering card for subsequent fits and also to increase flexibility of the metamorphs in this card by adding one or more CPs to the metamorphs. For the latter, we increase $N_m$ by the number of added CPs and insert these CPs as \texttt{``NEW"} in the ``\texttt{Xs Sm}" list.
The advantage of this practice is that insertion of the NEW CPs does not quantitatively change the metamorph saved from the previous fit, as the new CPs are placed exactly on the old modulator. The $\chi^2$ is exactly the same at the end of the previous fit and start of the new one, cf. an illustration in Fig.~\ref{fig:chi2_decr}. 

\subsection{The C wrapper}
\label{sec:CWrapper}
To facilitate the interface of \Fantomas with external codes, \texttt{FantoWrapper.cc} defines C functions invoking several core \Fantomas functionalities. These functions are used to link \Fantomas to the \xFitter code as described below. 

\begin{itemize}
    \item \texttt{readfantosteer\_()}:  This function will read the card file, \texttt{steering\_fantomas.txt}, by calling\\ \texttt{ReadCard()}, and initialize a \texttt{metamorphCollection} for the metamorphs specified in the steering card. In \xFitter, this function appears in the \texttt{atStart()} function in \texttt{Fantomas\_PdfParam.cc} to ensure that the collection is created once.
    \item \texttt{writefantosteer\_()}: The \Fantomas module produces an output steering card by calling \texttt{WriteCard()} and, optionally, other output files. In \xFitter, this function appears within \texttt{src/fcn.f} of the main \xFitter source code, and is designed to update the output file to represent the fit with the lowest $\chi^2$ encountered during the fitting process. This ensures that there are \Fantomas outputs if the fitting process is halted before completion.
    \item \texttt{updatefantopars\_(int \&ifl, double \*pars)}: updates parameters of the metamorph with the physi\-cal flavor \texttt{ifl} using the deltas at pointer \texttt{\*pars}, {\it i.e.}, it is a wrapper for the methods \texttt{UpdateParameters} and \texttt{UpdateMetamorphs} of \texttt{metamorphCollection}. \xFitter will vary the parameters with each iteration. This requires \xFitter to update the PDF values by calling \texttt{atIteration()}. The updated parameters are passed via the array of parameters from \xFitter, \texttt{pars}. The module keeps track of which \meta to use via the \texttt{flavor} argument.
    \item \texttt{fantopara\_(int \&ifl, double \&x)} returns the numerical value of the metamorph for a given physical flavor \texttt{ifl} and $x$ value. Within \xFitter, this function is called when \texttt{operator(x)} is called.
    \item  \texttt{fantoMellinMoment\_(int \&ifl, int \&MellinPower, int npts)} calculates the $(n+1)$-th Mellin moment, for $n=\texttt{MellinPower}$, physical flavor \texttt{ifl}, and \texttt{npts} integration points.
    \item \texttt{void getfantochi2\_(double\& fantochi2)} returns a contribution to the objective function ($\chi^2$) from the prior constraints on metamorphs. 
\end{itemize}

\subsection{The standalone \Fantomas implementation}
\label{sec:standalone}

The \texttt{standalone} directory of our Gitlab repository provides a lightweight implementation of the \Fantomas toolkit and simple usage examples. 
The standalone code illustrates the generation of \meta forms, updating the \meta parameters using the deltas, computation of Mellin moments, and updating the steering card. The standalone implementation does not require a minimization package; implementation in \xFitter will be described in Sec.~\ref{sec:xFitterImplementation}. 

\texttt{README.md} contains compilation instructions for the standalone examples. \texttt{FantomasExample\_Cpp.cc} and  \texttt{FantomasExample\_Wrapper.cc} respectively provide a sample \Fantomas run invoked directly in C++ and using the C wrapper from Sec.~\ref{sec:CWrapper}.

The C++ example first reads \texttt{steering\_fantomas.txt} to create an instance \texttt{metacol} of the \texttt{metamorph- Collection} class for  pion-like parametrizations: valence $xV(x)$, sea $xS(x)$, and gluon $xg(x)$. Recall that the metamorph collection stores all relevant information on the \meta functions for all flavors and accesses individual metamorphs according to their physical flavor \texttt{ifl}. It also provides methods to update all metamorphs at once or retrieve information from them.

 The executable \texttt{FantomasExample\_Cpp.x} created upon the compilation demonstrates these multiple features that are encountered during the typical minimization and will be described in the next subsection. There is no minimization involved in the standalone version, but the example imitates updates of the metamorph parameters in the following way.  \Fantomas reads the initial \meta parameters from \texttt{steering\_fantomas.txt}, while the external fitting program computes their changes (deltas). Thus, as seen from the fitting program, the fitted parameters are  the ``deltas", not the \meta parameters themselves. The example simulates the updates in deltas by randomly varying them. It then adjusts normalizations of some PDF flavors to satisfy the valence and momentum sum rules. At the end, the example writes the updated \meta parameters into  \texttt{steering\_fantomas\_out.txt}.   

The C++ example in \texttt{FantomasExample\_Cpp.cc} proceeds as follows.
\begin{itemize}
    \item It reads the steering card file \texttt{steering\_fantomas.txt}, described at length in Section~\ref{sec:steering_card}. If the printout of detailed diagnostics is requested (\texttt{VerbosityLevel=1}), it optionally prints out initial values of metamorph parameters, including \texttt{xPower}, \texttt{Sc[]}, and \texttt{Sm[]}. It then prints momentum fractions $\langle xf\rangle$, the values of the sum rules, and a table of PDF values vs. $x$.  
    \item The next part of the code emulates the \textsf{MINUIT} interface in which the delta parameters are defined for relevant \Fantomas parameters, \texttt{Sc[]} and \texttt{Sm[]}. The number of deltas should match the number of free parameters in all metamorphs, which requires manual editing of \texttt{FantomasExample\_Cpp.cc} and can be automated in the actual fitting program. 
    
    In the example, we initialize the deltas for the overall normalization \texttt{Sc[0]} to 1 and the rest of the deltas to zero. We then shift the deltas by a random amount between 0 and 1, to simulate the real minimization, in which parameters are varied by arbitrary amounts to adapt to the data. We update the metamorphs using the deltas and recompute some normalizations to satisfy the sum rules. Values of FIXed points remain unchanged.

    For instance, here is a part of the code that updates the gluon metamorph using a vector \texttt{deltag} containing the deltas.
\begin{python}
  // define the maximal orders of the metamorphs (number of CPs in steering_fantomas.txt minus 1)
  int Nmg = 3, NmS = 3, NmV = 3;
  // define the total number of parameters
  int ng = 6, nS = 7, nV = 7;                
  vector<double> deltag(ng + 1, 0.0);   // length = ng+1    
  // set normalizations to 1 for all flavors
  deltag[0] = 1;

  //Generate random shifts
  int seed=363432; 
  mt19937 gen(seed); // Mersenne Twister engine
  uniform_real_distribution<> dis(0.0, 1.0); // Range [0, 1)
  //Change deltas of the carrier...
  for (int i = 1; i < 3; i++){
    deltag[i] = dis(gen);
  }
  
  // ... and of the non-fixed CPs
  for (int i = 0; i < Nmg; i++)
    deltag[i+3] = dis(gen);

  //Update the metamorph
  metacol.UpdateParameters(iflg, deltag.data());
  //Compute and print the momentum fraction
  momg = metacol.MellinMoment(iflg,Mellinxf);
  cout << "<xg> = " << momg << endl;
  ...
  //Updated the gluon normalization, provided momentum fractions
  //for all three PDFs
  // Ag*<xg> + AS*<xS> + AV*<xV> = 1
  double Ag = (1 - momS - momV) / momg;
  deltag[0] = Ag;
  metacol.UpdateParameters(iflg, deltag.data());
\end{python}

The code prints out the current values of gluon parameters, deltas, and the momentum fractions before and after the metamorph normalizations:
\begin{python}
The deltas (changes in the parameters) are:
deltag={1.000000e+00, 9.303177e-02, 6.096957e-01, 8.542143e-01, 5.737977e-01, 2.180139e-01}
updated Scm[0][1] = 0.620432, deltas[1] = 0.093032
updated Scm[0][2] = 3.406296, deltas[2] = 0.609696
updated Scm[0][3] = 11.511414, deltas[3] = 0.854214
updated Scm[0][4] = 4.050728, deltas[4] = 0.573798
updated Scm[0][5] = 0.290044, deltas[5] = 0.218014
updated Scm[0][6] = 0.021282, deltas[6] = 0.000000
...
Updated momentum fractions before normalization:
<xg> = 0.14764369
<xS> = 0.28907919
<xV> = 0.39470286
<V> = 1.59060336
total momentum: 0.83142574
...
Momentum fractions after normalization:
<xg> = 0.21462756
<xS> = 0.28907919
<xV> = 0.49629325
<V> = 2.00000000
total momentum: 1.00000000
\end{python}
    \item At the end, the example saves \texttt{steering\_fantomas\_out.txt} with the updated metamorph parameters.
 \end{itemize}

Finally, \texttt{FantomasExample\_Wrapper.x} realizes the same example in C, rather than C++, using the wrapper. 

\subsection{The \xFitter implementation}
\label{sec:xFitterImplementation}
\subsubsection{Core files}
The \xFitter framework is reviewed in \ref{sec:xfit}. In the \xFitter environment, the \Fantomas toolkit implements a new parameterization of PDFs at the level of section ``3) Theory" of the flowchart of Fig.~\ref{fig:xfit}.  Our main files, including the \xFitter interface that descends from the 2020 pion fit  \cite{Novikov:2020snp}, are deployed in directory \texttt{pdfparams/Fantomas/}. In the top-level \texttt{CMake} files and main \xFitter source, we made a small number of changes tagged by the comment \texttt{\#lk} or \texttt{clk} and included via a preprocessor directive 
\texttt{ FANTOMAS\_XFITTER}. The flag that turns the directive on, \texttt{-DFANTOMAS\_XFITTER}, is enabled in the top-level \texttt{CMakeList.txt}. Finally, a new PDF flavor decomposition was required in \texttt{pdfdecomps/} for certain types of the pion fit.

In subdirectory \texttt{examples/Pion\_Fantomas/}, we provided control files to perform a sample pion fit with the \Fantomas parametrization. To run the sample fit, copy control files, including \texttt{steering\_fantomas.txt}, into the top \xFitter directory and run \texttt{bin/xfitter}. Upon successful completion within about 5 minutes, the fit returns the output files in \texttt{output/}. These files can be compared against the expected ones in\\ \texttt{examples/Pion\_Fantomas/output\_correct/}.

We will now elaborate on these augmentations. The main directory of the package, \texttt{pdfparams/Fantomas/}, contains the core source for \Fantomas (cf. Sec.~\ref{sec:Fantofiles}) as well as 
\begin{itemize}
    \item \texttt{CMakeLists.txt}: a CMake file, containing locations and names of the source code to be compiled. %
    \item \texttt{fantomas.cc} and \texttt{fantomas.h}: contain the \Fantomas C wrapper (Sec.~\ref{sec:CWrapper}) adapted for \xFitter.
    \item \texttt{Fantomas\_PdfParam.cc} and \texttt{Fantomas\_PDFParam.h}: realize the \xFitter interface to \Fantomas using the following functions:
    \begin{itemize}
        \item \texttt{atStart()} is executed when the parameterization is called the first time. It initializes all necessary class objects and variables.
        \item \texttt{atIteration()} is called every time the parameters are changed by \textsf{MINUIT}. It updates the modulator function with the new parameters.
        \item \texttt{operator(double x)} returns the numerical value of the metamorph function for a given $x$ value. 
        \item \texttt{moment(int N)} calculates the $N$-th moment; typically used to determine normalization according to the sum rules.
    \end{itemize}
\end{itemize}

When the $\delta$ parameters are changed by \xFitter, they are passed into our \meta functions only by calling the \texttt{atIteration()} method. Every such call recalculates the \meta functions using the updated parameters that were just provided. Similarly to the other parameterizations, \texttt{operator()} returns the overall PDF to the theory calculations within the \xFitter program: again, it is valid only after \texttt{atIteration()} has already updated the metamorphs following the change of parameters. Likewise, \texttt{moment()} passes the $N$-th moment from the internal calculation inside \Fantomas.

By the global nature of sum rules, they must be applied to the whole PDF function at once; hence some normalization constants are updated inside the interface file \texttt{Fantomas\_PdfParam.cc} and not in the \meta function.
In the example of the momentum fraction calculations given by Eq.~(\ref{eq:mmtSR}),  $A_S$ was taken to be a free parameter, determined by the minimization procedure, while $A_g$ is found by subtraction to the momentum sum rule. However, numerical instabilities may arise when fitting very small or zero gluon PDFs, affecting the final result \cite{Kotz:2023pbu,Kotz:2025lio}. To stabilize the fits with a small gluon, we had to consider $A_g$ as a free parameter and compute $A_S$ as a function of $A_V$ and $A_g$. For this scenario, we have introduced a new flavor decomposition for the pion PDF in 
\texttt{pdfdecomps\allowbreak /SU3\_(PositivePion\_)gluon}, effective at the level of the sum rules. It has to be specified in the configuration file \texttt{parameters.yaml} under \texttt{Decompositions}:
\begin{python}
Decompositions:                                                                          
  pion:                                                                                  
    class: SU3_PositivePion_gluon                                                        
    valence: v                                                                           
    sea: S                                                                               
    gluon: g   
\end{python}
\vspace{.5cm}

\subsubsection{$\chi^2$ penalties}
\label{sec:penal}
In the Bayesian paradigm, an \xFitter PDF fit maximizes the posterior probability $P(T|D)$, or equivalently it minimizes 
\begin{equation}
    \chi^2=-2\ln P(D|T) -2 \ln P(T) \equiv \chi^2_{\rm LL}+\chi^2_{\rm prior}
    \label{eq:objective}
\end{equation}
consisting of the log-likelihood $\chi^2_{\rm LL}$ and $\chi^2_{\rm prior}$ arising from the prior probability on the model. 
In realistic physics scenarios, the choice of the terms on the right-hand side comes with some freedom and plays an important role. 
The objective function $\chi^2$ is usually computed in the main source, {\it e.g.}, in a Fortran function \texttt{FCN} in \texttt{fcn.f} in \xFitter, 
where it should not be modified heavily. 
Our realization compiles \texttt{FCN} with a directive \texttt{FANTOMAS\_XFITTER} to compute $\chi^2_{\rm LL}$ and then add a  
 prior penalty $\chi^2_{\rm prior}$ on the metamorphs retrieved by $\texttt{call getfantochi2\_()}$.
The latter provides a C wrapper to a C++ method \texttt{metamorphCollection::Chi2prior()} 
defined in a short dedicated file \texttt{metamorphCollectionPrior.cc}. When the $\chi^2$ value is reduced, \texttt{FCN} saves
\texttt{steering\_fanto\allowbreak mas\_out.txt}, containing the latest metamorph parameters, and
\texttt{metamorphCollectionPrior.cc}, containing the prior that may change between the fits. Both files are saved in \texttt{output/}, so that
the lowest attained $\chi^2$ can be reproduced if the fit is interrupted or repeated. 

The included \texttt{metamorphCollectionPrior.cc} provides a basic realization of \texttt{Chi2prior()} consisting of several blocks for common types of penalties that may be imposed, including
\begin{itemize}
    \item lasso and ridge penalties on CP values,  $\sum_{i=1}^{N_m+1}|P_i|$ and $\sum_{i=1}^{N_m+1}P_i^2$, respectively; these can be activated to constrain the total deviation of the modulator from unity, introducing rigidity to the modulator form.
    \item activation functions constraining the growth of a PDF-dependent quantity $Y$, including the (rectified linear unit) activation function ${\rm ReLU}(Y)\equiv Y\theta(Y)={\rm max}(0.0, Y)$, and logarithmic activation function $\left|\log(1+ |Y|)\right|$. $Y$ can typically be a metamorph at some $x$, metamorph parameter, or Bernstein coefficient $C$. 
\end{itemize}

By default, the blocks of penalties are commented out and can be uncommented as needed to assemble 
\begin{equation}
    \chi^2_{\rm prior} = \sum_j\, w_j \chi^2_j,
\label{eq:chi2model}
\end{equation}
where $\chi^2_j$ is the penalty from the $j$-th block included with weight $w_j$ that is also specified inside \texttt{Chi2prior()}. 
Such priors may, {\it e.g.}, guide the gluon's fall-off moderately at large $x$, the modulator function to effectively subside to the carrier, or ensure integrability, {\it e.g.}, at very small $x$ values, through limiting the allowed values of the $B$ parameters.

On the last point of \textbf{integrability}, the flexibility of the \meta functions may lead to non-convergence of lowest Mellin moments, which has to be addressed at the level of individual PDF parametrization classes and not in the \xFitter main code. Each parameterization class within \xFitter provides an integration method specifically for the said class. Integrability can be checked, {\it e.g.}, when enforcing the sum rules. Class \texttt{metamorph} does not enforce integrability on the \meta and instead issues a warning message when the integrand is potentially non-integrable. The provided \meta functions are regular except in the limits $x\to 0$ and $x\to 1$, and hence integrability can be checked based on the asymptotic powers $B_a$, $C_a$ of the carrier for $xf(x)$ and the requested power $n$ of the Mellin moment $\langle x^nf\rangle$. [Here $B_a$, $C_a$ are the current, not initial values.] \texttt{metamorph::GetMellinMoment()} warns that the $n$-th moment is non-integrable if either $B_a + n \leq -1 $ or $C_a  \leq -1$. If either combination is between -1 and -0.93, it warns about slow convergence of the moment. 

If the warning occurs, such solutions can be disfavored by including a prior contribution, $w\ {\rm ReLU}(-B_a - n - 1) $ or $w\ {\rm ReLU}(-C_a - 1) $.
Analogous prior constraints can be imposed directly on the PDF values, {\it e.g.}, to enforce \textbf{positivity}. 
We encourage the use of such priors over that of the bounds on parameters provided by, {\it e.g.} \textsf{MINUIT}, to control the prior constraints.

\subsubsection{More on inputs and outputs \label{sec:xFitterIO}}
The \Fantomas steering card, \texttt{steering\_fantomas.txt}, must be placed in the top \xFitter directory. A few other standard files of the \xFitter package are pertinent to the \Fantomas-based fits:
\begin{itemize}
    \item \texttt{parameters.yaml} is a steering file that defines physical flavor IDs and delta parameters passed into metamorphs, as it also serves as the control file for the \textsf{MINUIT} minimization. The flavor ID and parameter assigments in this file must match the ones defined within \texttt{steering\_fantomas.txt}. \ref{sec:xfit} summarizes the contents of \texttt{parameters.yaml} in the case of \Fantomas fits, and it also lists other relevant input files. 
    \item \texttt{output/MINUIT.out.txt} reports the status of the \textsf{MINUIT} minimization, including the final values of the delta parameters. 
     \item \texttt{output/xFitterPi/} is an existing \xFitter output directory containing produced LHAPDF grids\\ (\texttt{xFitterPi\_XXXX.dat} and \texttt{xFitterPi.info}).\footnote{For the users who are not familiar with the LHAPDF format of \url{https://www.lhapdf.org/}, a \texttt{Mathematica} package that reads the grids, ManeParse, is available here: \url{https://ncteq.hepforge.org/mma/index.html}.}
\end{itemize}

Finally, we note that a fit can simultaneously deploy \Fantomas and non-\Fantomas parametrizations, depending on the specific hadron and flavor. For example, in the pion fit we simultaneously use the \Fantomas parametrizations for pion PDFs and LHAPDF parametrizations for nuclear PDFs. In such situation, the code runs additional checks to initialize both \meta and non-\meta functions.

\section{Practical example: pion PDFs}\label{sec:FantoPDFs}

\begin{figure}[t]
     \centering
     \includegraphics[width=0.65\linewidth]{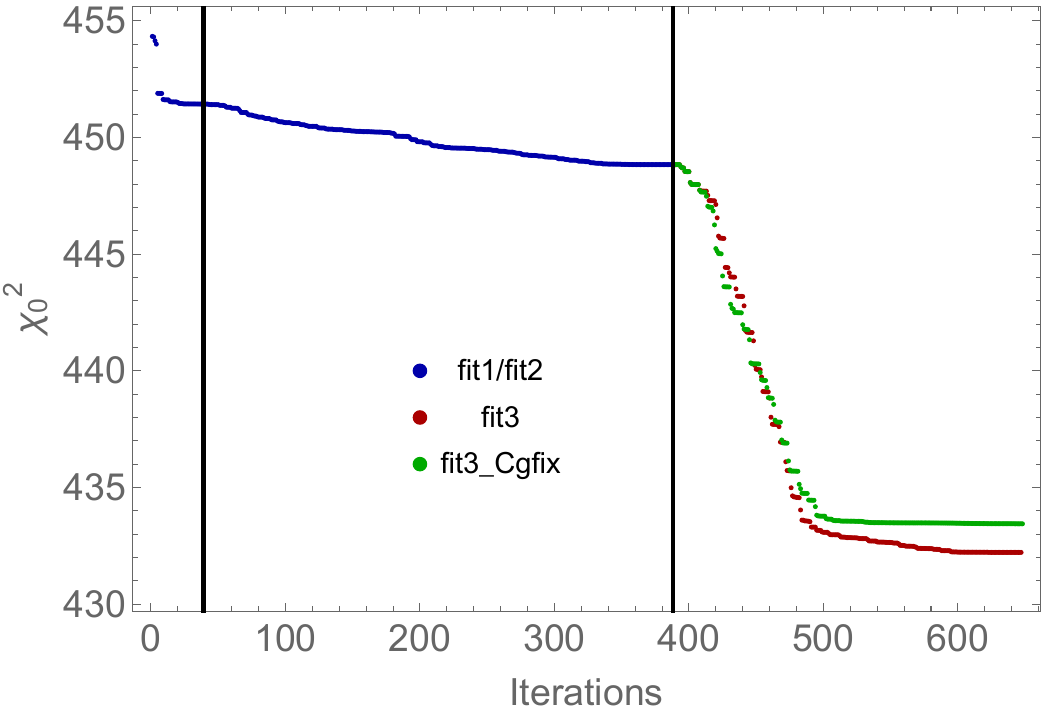}
\caption{ Illustration of $\chi_0^2$, the lowest $\chi^2$ value when an improvement is detected in the fitting process, which decreases when adding CPs. The number of iterations is accumulated for the three fits (fit1, fit2, and either fit3 or fit3\_Cgfix) corresponding to the addition of CPs, as the starting point for each higher degree polynomial comes from the best fit of the previous. See text. }
\label{fig:chi2_decr}
\end{figure}

 To illustrate the possibilities at hand when using the \Fantomas toolkit, we go back to the fits of PDFs for charged pions~\cite{Kotz:2023pbu,Kotz:2025lio}, consisting of the valence $xV(x)$ [$u-\bar u=\bar d-d=V/2$ in $\pi^+$], SU(3)-symmetric sea $xS(x)$, and gluon $xg(x)$ distributions.
{\bf The parametrization in Eq.~(\ref{eq:metamorphbasis}), together with Eq.~(\ref{eq:modulator}), is designed to  progressively increase the flexibility of the PDFs, by first realizing the asymptotic behaviors with the carrier and adding control points providing modulation at mid-$x$ values. The $\chi^2$ decreases without interruptions under this procedure, as illustrated in an example in  Fig.~\ref{fig:chi2_decr}.} 
This section is centered around this example. It shows how to construct a family of \meta parametrization starting from the carriers, then incrementing the flexibility of functional forms while introducing prior penalties to moderate unconstrained behavior of PDFs if the data are limited.

\begin{wrapfigure}{BTH}{0.35\textwidth}
     \centering
     \includegraphics[width=\linewidth]{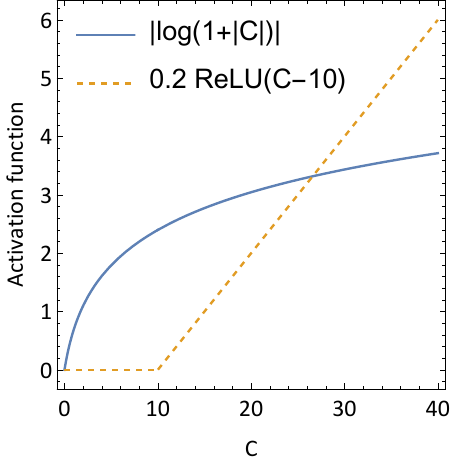}
\vspace{-25pt}
\caption{Activation functions listed in Table~\ref{tab:4penalties}, with indicated weights. 
}
\label{fig:activation_function}
\end{wrapfigure}

The data sets used for the pion PDF fit include three different processes -- pion-induced Drell-Yan, prompt photon and leading-neutron DIS -- for a total of 408 data points, with ranges in $x_{\pi} \in[\approx\! 10^{-3}, \approx\! 1]$ and $Q^2\in [\approx\! 7, \approx\! 250]$ GeV. 
Previous studies (Refs.~\cite{Barry:2018ort, Kotz:2023pbu}) have already emphasized the sparse coverage in the low- and mid-$x$ regions. Even with the addition of the leading-neutron data, the mid-$x$, small-$Q^2$  region remains unpopulated, effectively making it an extrapolation region.  This paucity of the pion data should be kept in mind, as it requires to introduce prior constraints in several cases to guarantee convergence of the fits.

In some presented examples, we introduce the penalties defined in Sec.~\ref{sec:penal} and with the activation functions and weights listed in Table~\ref{tab:4penalties}. The \texttt{AbsLog} and \texttt{ReLU I} penalties both favor Bernstein coefficients $|c_l|$  of order $\mathcal{O}(1)$; in the examples, we impose such a penalty on all three flavors. 
The rest of the penalties disfavor some unphysical solutions for PDFs that are allowed due to the scarcity of the pion data. \texttt{ReLU II} favors a positive valence PDF  at mid- to large-$x$ values, while \texttt{ReLU III} disfavors a gluon PDF that falls off too rapidly, as $(1-x)^{C_g}$ with $C_g>10$, at $x\to 1$. Penalties II and III call the PDFs at a given $x$ and $Q_0$. 

Fig.~\ref{fig:activation_function} shows the \texttt{AbsLog} and \texttt{ReLU} penalties with $w=1$ and 0.2. 
While the $|\ln(1+|C|)|$ (\texttt{AbsLog}) penalty ``softly" turns on and disfavors large $C$ values, the $\texttt{ReLU}(C-C_0)$ is more abrupt in that it is absent for $C<C_0$ and rapidly grows for $C>C_0$. Therefore, the \texttt{ReLU} penalty does not affect the fit at all as long as $C$ is moderate.  Such priors may, {\it e.g.}, guide the gluon to fall off moderately at large~$x$, force the modulator function to effectively subside relative to the carrier,  or ensure integrability as $x \to 0$ by limiting the allowed values of the $B$ parameters. Other priors can be imposed, in addition to the ones listed, offering some advantages over the bounds on parameters provided, {\it e.g.}, by \textsf{MINUIT}.

\begin{table}[h!]
\centering
\renewcommand{\arraystretch}{1.5} %
\begin{tabular}{|c|c|c|}
\hline
\textbf{Penalty name} & \textbf{Activation function} & \textbf{Weight} $w$ \\ \hline
\texttt{AbsLog}          & $|\log(1+|c_l|)|$      & 1   \\ \hline
\texttt{ReLU I}       & $\mathrm{ReLU}\left[|c_{l}|-10\right]$      & 1/($N_m+1)$  \\ \hline
\texttt{ReLU II}      & $\mathrm{ReLU}\left[\bar u(x, Q_0)- u(x, Q_0)\right], x=0.1 $     & 50     \\ \hline
\texttt{ReLU III}     & $\mathrm{ReLU}\left[ C_g-10  \right]$ & 50  \\ \hline
\end{tabular}
\caption{Penalties imposed for $\pi^-$ PDF determination.} 
\label{tab:4penalties}
\end{table}

\begin{figure}[p]
\center
 \begin{python}
# gluon -- metamorph 0                                                                   
# ifl  Nm  MappingMode  xPower  Sc(0)  Sc(1)  Sc(2)                                             
    0   1    0            1.    0.42    -0.37   2.83                                                     
# Xs  Sm                                                                         
0.001 FIX                                                                                
0.8   FIX                                                                                
# sea  -- metamorph 1                                                                    
# ifl  Nm  MappingMode  xPower  Sc(0)  Sc(1)  Sc(2)                                             
    1   1    0           1.      9.79    0.73    8.19                                                     
# Xs  Sm                                                                          
0.001 FIX                                                                                
0.8   FIX                                                                                
# valence  -- metamorph 2                                                                
# ifl  Nm  MappingMode  xPower  Sc(0)  Sc(1))  Sc(2)                                             
    2   1    0            1.     2.54    0.74    0.95                                                     
# Xs  Sm                                                                          
0.001 FIX                                                                                
0.8   FIX 
 \end{python}
(a)
\begin{python}
# gluon -- metamorph 0
# ifl  Nm  MappingMode  xPower  Sc(0)  Sc(1)  Sc(2)                         
    0   2    0            1.    0.42    -0.37   2.83
# Xs  Sm  
0.001	FIX
0.4	NEW
0.8	FIX
# sea  -- metamorph 1
# ifl  Nm  MappingMode  xPower  Sc(0)  Sc(1)  Sc(2)  
    1   2    0           1.      9.79    0.73    8.19 
# Xs  Sm 
0.001	FIX
0.4	NEW
0.8	FIX
# valence  -- metamorph 2
# ifl  Nm  MappingMode  xPower  Sc(0)  Sc(1))  Sc(2)                             
    2   2    0            1.     2.54    0.74    0.95
# Xs  Sm 
0.001	FIX
0.4	NEW
0.8	FIX
\end{python}
(b)
 \caption{Initial steering cards for (a) the simplest $N_m=1$ fit with 2 FIXed CPs and (b)
 an extended fit with 2 FIXed, 1 FREE CPs per flavor. \label{fig:2fix}}
\end{figure}

\subsection{Building up a \meta}
\label{sec:0thfit}

\textbf{A carrier-like metamorph.} When building the parametrizations from scratch, the simplest metamorph is given by a carrier function multiplied by an $N_m=1$ modulator $F^{mod}=1$ with two fixed CPs set at the lower and upper ends,
denoted by $x_{\rm min}$ and $x_{\rm max}$, of the $x$ region covered by the experimental data. Recall from Eq.~(\ref{eq:stretch}) that the modulator's argument is given by a stretching function $y(x)=g(x,n)^{\alpha_x}$, where $g(x,n)$ freezes the $x$ dependence below $x_{min}$ as well as above $x_{max}$. If the lowest (highest) CP is FIXed, its $x$ value sets the value $x_{min}$ ($x_{max}$). With the two fixed CPs, $F_{\rm mod}=1$. Hence, the $N_m+1=2$ control points translate into $c_{0,1}=0$, leading to an effective $N_m^{\rm eff}=0$ \meta. 

\textbf{One free control point.} Figure~\ref{fig:2fix}(a) shows a sample steering card for such fit, called ``fit1" in the following. Contrast this with another steering card for ``fit2" in Fig.~\ref{fig:2fix}(b), in which we elevated degree $N_m$ to 2 (by one unit) in lines 3, 10, and 17, and accordingly added NEW CPs in lines 6, 13, and 20. The NEW CPs are initialized with $P_i=0$, so that they lie exactly on the parent metamorph. 
In the output steering card, the lines for these CPs contain their numerical $P_i$ value instead of ``NEW".

Since the FIXed CPs do not change their values, in general they do not require assignment of a free parameter for the respective deltas. In the \xFitter realization, on the other hand, each CP requires an assigment of a delta parameter in \texttt{parameters.yaml}, but the values for the FIXED CP's can be left at zeros and will not change.  We illustrate the respective record for the sea distribution in ``fit2":
        \begin{python}
        Parameters: 
        ...
            delDs: 0.                                                                              
            delEs: [ 0., 0.1]                                                                      
            delFs: 0. 
        ...
        Parameterisations:                                                                       
        ...
          S:                                                                                     
           class: Fantomas                                                                      
           parameters: [As,delBs,delCs,delDs,delEs,delFs,sifl] 
        ...
        \end{python}
\texttt{delDs} and \texttt{delFs} stand for the free parameters for deltas of the CPs at $x=0.001$ and $x=0.8$. Their values (here set to 0. to comply with the \xFitter internal convention) are ignored inside the \meta. In the other fitting programs, specifying deltas for these FIXed CPs would be redundant. The delta parameter \text{delEs} for the FREE CP at $x=0.4$, however, has the initial value and initial step of 0. and 0.1, respectively. It corresponds to an active \texttt{MINUIT} parameter. 

\textbf{Two free control points.} 
Once ``fit2" converged, we add a NEW free CP to $xV$ and to $xS$, raising $N_m$ to 3 for these flavors. We keep 1 FREE CP ($N_m=2$) for the gluon, the least constrained PDF flavor. That is, our delta parameters for the CP's of $xS$ are now \texttt{delCs,delDs,delEs,delFs,delGs}, with \texttt{delCs} and \texttt{delGs} for the two FIXed CPs now fixed at zero values.
For a carrier-only setting ``fit1",  each of the three flavor combinations had three free parameters, for a total of 9 parameters, two of which are determined by the valence sum rule and momentum sum rule. 
With five free CP's now at play, we have a total of 12 free parameters, which leads to flat directions in the PDF parameter space. We must constrain these directions either by fixing some parameters or adding prior penalties. In the following, we show results of such two fits, ``fit3" and ``fit3Cgfix". 

\begingroup
\setlength{\tabcolsep}{10pt} %
\renewcommand{\arraystretch}{1.5} %
\begin{table}[t]
    \centering
    \begin{tabular}{|l|c|c|c|c|c|c|c|}
    \hline 
        Fit name                  & Free CPs & $N_{par}$ & Penalty & $\chi^2$ & $\langle xg \rangle$ & $\langle xS \rangle$ & $\langle xV \rangle$\\
        \hline \hline
        \multicolumn{8}{|c|}{Main sequence}\\ \hline
        fit1                     & \{0,0,0\} &    7      & \texttt{ReLU}     & 451.44 & 0.269 & 0.180 & 0.551\\ \hline
        fit2                     & \{1,1,1\} &   10      & \texttt{ReLU}     & 448.84 & 0.257 & 0.187 & 0.556\\ \hline
        fit3 (no convergence)    & \{1,2,2\} &   12      & \texttt{ReLU}     & 432.61 & 0.120 & 0.448 & 0.432\\ 
       fit3Cgfix (fixed $C_g$)  & \{1,2,2\}  &   12      & \texttt{ReLU}     & 433.43 & 0.130 & 0.427 & 0.443\\
        \hline \hline
        \multicolumn{8}{|c|}{Alternative fits}\\ \hline
       fit2alt                  & \{1,1,1\} &   10      & \texttt{AbsLog}   & 451.18 & 0.269 & 0.180 & 0.551\\
       fit1g                    & \{0,0,0\} &    7      & \texttt{AbsLog}     & 451.44 & 0.269 & 0.180 & 0.551\\
        fit2g                    & \{1,1,1\} &   10      & \texttt{AbsLog}   & 451.18 & 0.269 & 0.180 & 0.551\\
        \hline 
    \end{tabular}
    \caption{The NLO pion fits discussed in this section: name of the fit; numbers of free CPs in $xg$, $xS$, and $xV$; number of free parameters $N_{par}$; the penalty type; best-fit $\chi^2$ values, and best-fit momentum fractions. The first four (last two) fits take the sea normalization $A_S$ (the gluon normalization $A_g$) to be a free parameter. The $\chi^2$ function includes both the log-likelihood and penalty contribution, as in Eq.~(\ref{eq:objective}). For two versions of fit2, small changes in $\chi^2$ and momentum fractions reflect the differences between the \texttt{AbsLog} and \texttt{ReLU} penalties on Bernstein coefficients $c_l$. 
    }
    \label{tab:fantochi2tab}
\end{table}
\endgroup

\subsection{Exploring the pion fits}
\label{sec:PionFits}
\textbf{Dependence of $\chi^2$ on the polynomial degree.} The four just described fits, created by adding the free CPs as described, are listed as a ``main sequence" in the upper half of Table~\ref{tab:fantochi2tab} summarizing their settings. In this sequence, we uniformly apply the $\texttt{ReLU}$ priors from Table~\ref{tab:fantochi2tab} to guarantee the preferred physical behavior\footnote{Penalty \texttt{ReLU I} is not activated in the carrier-only fit1, as all $c_l$'s are zero.} -- but other solutions can be obtained by choosing the other priors or eliminating them entirely. 

\begin{figure*}[t]
     \centering
     \includegraphics[width=0.49\linewidth]{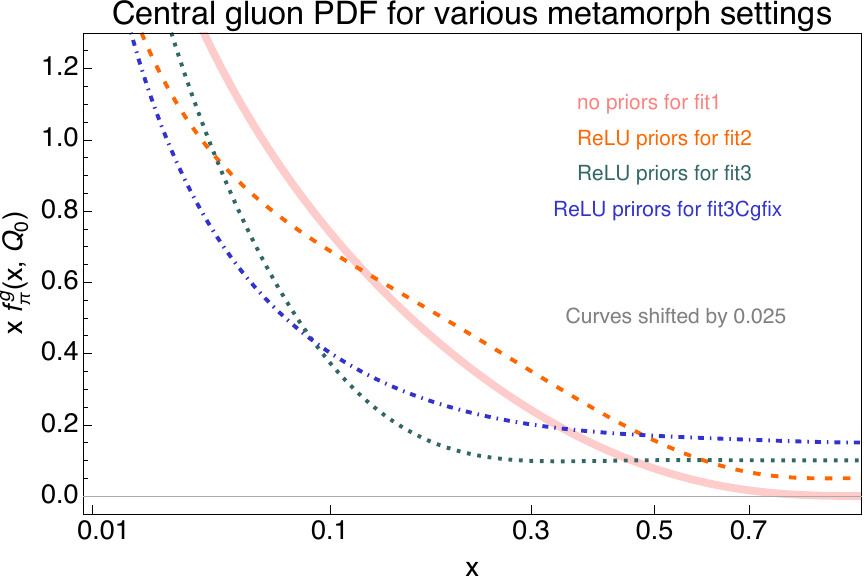}
     \includegraphics[width=0.49\linewidth]{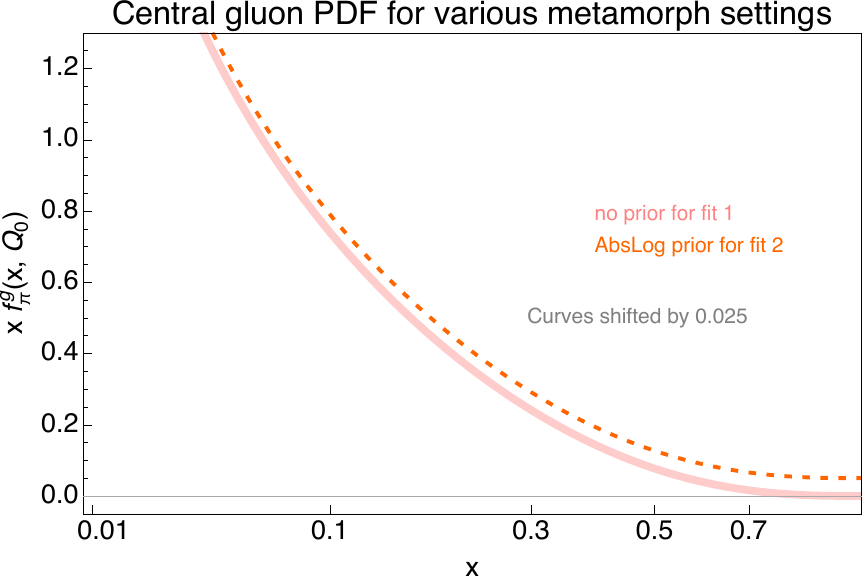}
\caption{Central values for the gluon PDF of the pion using \texttt{ReLU} priors (left) and  \texttt{AbsLog} priors  (right), given at $Q_0$, for the main sequence of fits Table~\ref{tab:fantochi2tab}. The consecutive curves have been shifted up by multiples of 0.025 for clarity. For fit1, neither of the described penalties have an effect.}
\label{fig:shift_central_nominal}
\end{figure*}

Figure~\ref{fig:chi2_decr} shows the dependence of  $\chi^2$ on the number of minimization iterations. The left Fig.~\ref{fig:shift_central_nominal} illustrates the variety of the PDF solutions obtained by increasing the polynomial degree, here visualized with the resulting gluon PDFs. Table~\ref{tab:fantochi2tab} gives the resulting $\chi^2$ and momentum fractions of the four fits.

Turning first to the $\chi^2$ in Fig.~\ref{fig:chi2_decr},  
by adding the CPs to the solution of a converged fit, and hence increasing the degree of the polynomial of the modulator, we expect to see a decrease in the $\chi^2$ value. In fact, the decrease happens with any polynomial parametrization, but only after all coefficients were refitted. With the common procedure that directly fits the polynomial (Bernstein) coefficients, the $\chi^2$ initially increases after changing the polynomial's degree. This $\chi^2$ increase does not happen with the \meta parametrizations: when a new fit starts, addition of free CPs does not modify the progenitor \meta function of a lower degree, and all Bernstein coefficients are recomputed accordingly. Hence, with the described procedure to increase the degree of the polynomial, the evaluated $\chi^2$ initiates at an already low value found with the progenitor parametrization: we avoid cumbersome and redundant evaluations when the parametrization is changed. 

This feature is illustrated in Fig.~\ref{fig:chi2_decr}, where $\chi^2$ remains continuous when we switch from fit1 to fit2 and then to fit3. The first iterations represent the optimization corresponding to fit1. 
The vertical lines show the initialization of the following step, {\it i.e.}, fit2 or fit3. The $\chi^2$ starts at the exact same minimum value as that found in the last iteration of previous step of the procedure. 

Two versions of fit3 are illustrated, representing a baseline and a modified version of the same settings. For the sea and the valence, fit3 uses four CPs for each, with two CP's FIXed at the end points. The gluon PDF is kept at a lower order with only one free CP. With the fit settings in Table~\ref{tab:4penalties}, fit3 does not converge -- the data do not allow for two extra parameters to be determined. It is illustrated by the red dots in Fig.~\ref{fig:chi2_decr}. When fixing the large-$x$ behavior of the gluon to the best-fit value of the previous step ($\delta C_g=0$), fit3Cgfix converges and the $\chi^2$ reaches a plateau at a slightly higher value than fit3, shown by the green dots.

Figure~\ref{fig:shift_central_nominal} shows that the shape of the gluon PDF changes significantly as we progress through the sequence of four fits. Clearly, the order of the polynomial approximation matters for the PDF's behavior. In addition, the fits with at least one free CP (fit2 and fit3) are sensitive to the choice of the prior, as we illustrate by comparing the left and right panels of Fig.~\ref{fig:shift_central_nominal}, obtained with the \texttt{ReLU} and \texttt{AbsLog} priors in Table~\ref{tab:4penalties}, respectively.

For fit1, neither of the considered penalties has an effect even if we were to impose them. They are not imposed as a result:
the fit1 curves are identical in the left and right panels.

The form of the penalty is more important for the rest of the fits. For fit 2, in addition to the version with the \texttt{ReLU} prior penalty in the left panel, we alternatively present a version constrained by the \texttt{AbsLog} prior with a weight of $w=1$.
The resulting gluon is displayed in the right panel of Fig.~\ref{fig:shift_central_nominal}, and the $\chi^2$ as well as momentum fractions are listed in the first row of ``Alternative fits", labeled as "fit2alt", in Table~\ref{tab:fantochi2tab}. 
The \texttt{AbsLog} penalty has a smoothly increasing objective function, see Fig.~\ref{fig:activation_function}, which mildly disfavors even non-zero Bernstein coefficients, $|c_l|$ of O(1). Consequently, the fit2alt curve in the right panel must be very close to the carrier-only fit1 curve, as its allowed $|c_l|$ are small; the $\chi^2$ of fit2alt in Table~\ref{tab:fantochi2tab} is only a tad below that for fit1. 

On the other hand, going back to the \texttt{ReLU} penalties in the left panel, the Bernstein coefficients are allowed to grow up to $|c_l|\sim 10$: the fit2 gluon is significantly different from the fit1 gluon, despite the fact that $\chi^2$ values are practically the same for fit1 and fit2. The direct constraint on the gluon is weak. We rather observe the interplay between different flavors, with the rectilinear activation impeding the sea PDF when one of the $c_l$ of the sea saturates the bound of ${\cal O}(10)$, hence modifying the $\chi^2$. As the sea is intertwined with the gluon, both PDFs are affected and result in two different fit1 and fit2, as seen from comparing the fit2 curves in the two panels of Fig.~\ref{fig:shift_central_nominal}.

The \texttt{ReLU} penalties are consequential for fit3 as well. In the left panel, the distinct behaviors of fit3 and fit3Cgfix are mostly due to the \texttt{ReLU III} penalty on the suppression power $C_g$ in the large-$x$ power-law falloff $g(x,Q_0)\propto (1-x)^{C_g}$. While $C_g$ can be large in fit3, strongly suppressing $g(x,Q_0)$ across most of the $x$ range, the \texttt{ReLU III} penalty disfavors such suppression of fix3Cgfix, and hence it's the prior penalty, not the data, that determines the slower falloff of the fit3Cgfix gluon at $x\to 1$.

We have also explored the cases for which $A_S$ is determined from the momentum sum rule instead of $A_g$, that is, for which $A_g$ is a free parameter. 
Recall that fitting $A_g$ is preferable when the gluon PDF is small, as is allowed by the contemporary pion data. The last two rows of Table~\ref{tab:fantochi2tab} lists settings of two such fits, dubbed ``fit1g" and ``fit2g" -- the counterparts of ``fit1" and ``fit2alt". \xFitter implements the free-$A_g$ fits with the help of the \texttt{pdfdecomps/SU3\_(PositivePion\_)gluon} flavor decompositions included in our \xFitter branch. We expect that the free-$A_S$ and free-$A_g$ fits to be equivalent when they converge well -- and indeed we see such equivalence between the ``fit1" and ``fit1g", and ``fit2alt" and ``fit2g" rows of the Table.

\subsection{Channeling flexibility}
\label{sec:flexibility}

In this section, we continue to showcase the properties of metamoprhs when generating diverse functional forms, now exploring dependence on such inputs as $\alpha_x$, $N_m$, or the positions of fixed and free CPs in Eq.~(\ref{eq:modulator}), still using the pion fits as the example. 
We take a variation on fit1 and fit2 with even weaker \texttt{AbsLog} penalties with $w=0.01$. This variation, labeled ``Nominal" from now on, has one free CP at $x=0.4$ for the sea and the valence (no free CP for the gluon), plus two FIXed CPs at $x=0.001, 0.8$ for every flavor. With our choice of weak penalties, the fit is very similar to the carrier-like result [fit1] shown in the right panel of Fig.~\ref{fig:shift_central_nominal}.

By modifying this nominal result, we will explore three features, namely, the invariance under a relocation of free CPs, variation under a relocation of fixed CPs, and a change due to the stretching power, $\alpha_x$. 

\begingroup
\setlength{\tabcolsep}{10pt} %
\renewcommand{\arraystretch}{1.5} %
\begin{table}[tb]
    \begin{tabular}{|l|c|c|c|c|l|}
    \hline 
        Fit name           &$\chi^2$ & $\langle xg \rangle$ & $\langle xS \rangle$ & $\langle xV \rangle$ & Description\\
        \hline
        Nominal            & 446.99 & 0.228 & 0.258 & 0.515 & {\small \{0,1,1\} free CPs, 9 parameters, $\alpha_x=1$}\\
        Change in free CPs  & 446.99 & 0.228 & 0.258 & 0.515 & {\small Free CPs moved from $x=0.4$ to 0.3}\\
        Change in fixed CPs & 445.94 & 0.191 & 0.307 & 0.503 & {\small Fixed CPs moved, $x=\{10^{-3}, 0.8\} \to \{0.1,0.6\}$ } \\
         $\alpha_x=0.3$   & 444.41 & 0.090 & 0.482 & 0.428 & {\small $\alpha_x$ changed to 0.3}\\
        \hline
    \end{tabular}
    \caption{$\chi^2$ values and momentum fractions for a family of fits obtained from the ``nominal" baseline by changing the various settings discussed in the text.}
    \label{tab:fantochi2tabComp}
\end{table}
\endgroup

\begin{figure}[t]
    \centering
    \includegraphics[width=0.45\linewidth]{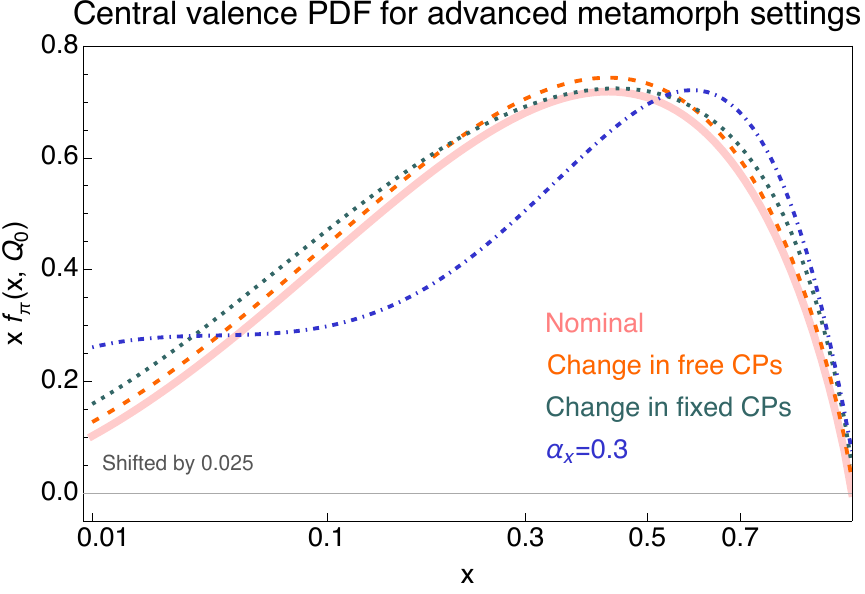}
    \includegraphics[width=0.45\linewidth]{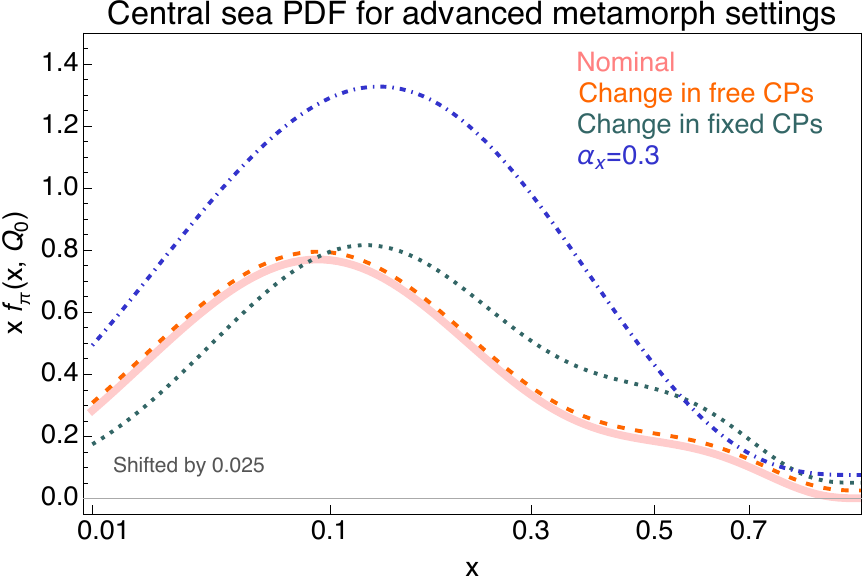}
    \includegraphics[width=0.45\linewidth]{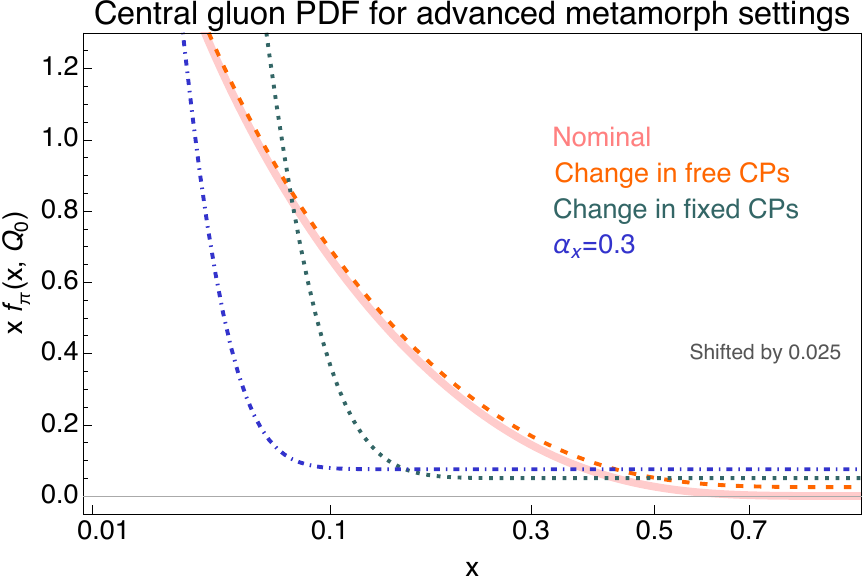}
\caption{PDFs from fits in Table~\ref{tab:fantochi2tabComp} illustrating the impact of \meta settings: nominal fit (solid pink), change in all free control points (dotted orange), change in all fixed control points (dashed dark green) and change in $\alpha_x$ (dot-dashed dark blue). See text. The curves are shifted incrementally by $0.025$ in the order they appear in the legend. }
    \label{fig:2fix2freePDFs}
\end{figure}

Figure~\ref{fig:2fix2freePDFs} shows the outcomes, for all three flavor combinations, of those three variations available within the \Fantomas methodology. The \meta labeled ``Change in free CPs" were obtained modifying the position of the free CP for the sea and the valence from $x=0.4$ to $x=0.3$. The corresponding \meta PDFs (dashed orange curves) did not change compared to the nominal ones (pink solid curves). In these plots, the consecutive curves are shifted upward by 0.025 to distinguish coinciding PDFs. 

Hence, given some $N_m$, positions of free CPs can be moved around without changing the best-fit polynomial solution. The only restriction is that free CPs cannot be too close so as not to introduce numerical instabilities, cf. \ref{sec:num}. The invariance of the best-polynomial with respect to the placement of free CPs is mathematically rigorous. Namely, since the best-fit polynomial of a given degree $N_m$ and the given number of free CPs is unique according to the unisolvence theorem, moving the free CPs around does not change this polynomial. Consequently, the PDFs labeled as ``Change in free CPs" in Fig.~\ref{fig:2fix2freePDFs} are equivalent to the nominal ones.

The metamorphs labeled ``Change in fixed CPs" were obtained by modifying the positions of the fixed CPs of all three combinations from $x=0.001, 0.8$ in  ``Nominal" to $x=0.1, 0.6$. This produces noticeable variations in the \meta (dotted dark-green curves), as expected.

The metamorphs labeled ``$\alpha_x=0.3$" were obtained by modifying the stretching exponent for all three combinations from $\alpha_x=1$ in ``Nominal" to $\alpha_x=0.3$. Again, there are noticeable variations in the \meta (dot-dashed dark-blue curves). \emph{The choice of the stretching function $y(x)=x^{\alpha_x}$ thus is a meaningful control to modify the \meta, together with the choice of $N_m$ and placement of fixed CPs.}

\textbf{How many free parameters?} As we increase the flexibility of the parametrization by adding parameters, we eventually reach a low $\chi^2$ value for which the fit effectively stops converging. The unisolvence theorem again explains why this happens -- see the discussion in \cite{Courtoy:2020fex}. Namely, as the number of parameters increases, we transit from a single polynomial solution achieving the $\chi^2$ minimum, when the data and prior constraints are constraining enough, to infinitely many solutions with comparably good $\chi^2$. 
In some cases, non-convergence may happen, as illustrated with ``fit3." As shown, one may recover the regime with a global minimum of $\chi^2$ by fixing select parameters or adding prior constraints. The representative exploration of a variety of solutions, including higher degree polynomials that may introduce redundancy, remains necessary as a part of estimation of the epistemic uncertainty, since the solutions with about the same $\chi^2$ may be equally acceptable. %
The \Fantomas toolkit enables robustness when exploring a range of solutions rather than locking into a potentially over-constrained space. When quantifying the resulting uncertainty, the value of the best $\chi^2$ alone does not constitute a sufficient criterion in favor of a model/fit. In the next Section, we will discuss how various models can be combined to form a PDF ensemble that accounts for the diversity of such parametrizations.

\section{Combination of PDF fits \label{sec:combination}}

In addition to the central fit corresponding to the global minimum of $\chi^2$, a global QCD analysis in the Hessian formalism typically provides an ensemble of eigenvector PDF sets to quantify experimental and other sources of the uncertainty. In \xFitter, the default criterion to define the $1 \sigma$ experimental uncertainty for a given PDF parametrization corresponds to a cut in the chi square increase of $\Delta \chi^2=1$ above the best-fit value.

However, the final PDF uncertainty must account for the span of acceptable PDF solutions that can be determined using a number of criteria. The most common procedure to combine the PDF solutions with distinct functional forms is based on model averaging using the $\chi^2$ or a related objective function of the type given in Eq.~(\ref{eq:objective}). This is the simplest approach, however, associated with the \textit{weak} goodness-of-fit criterion \cite{Kovarik:2019xvh} quantifying the \textit{average} agreement of the theoretical model and data. A model satisfying the weak criterion may still fail in some ways, {\it e.g.}, by disagreeing with subsets of data beyond the level allowed by random fluctuations of data. The \textit{strong} goodness-of-fit criteria \cite{Kovarik:2019xvh} are more comprehensive ones; they quantify more detailed comparisons of theory and data, but generally are more difficult to implement. Bayesian model averaging must also generally account for dependence on the priors and can be done, {\it e.g.}, using the methods adopted in lattice QCD \cite{Jay:2020jkz} and recently explored for toy PDFs \cite{Yan:2024yir}. In such Bayesian combination, information criteria penalize models with too many parameters. Finally, the combination must account, even reward, for including maximally dissimilar PDF solutions that provide comparable agreement with the fitted data.

For example, in Refs.~\cite{Kotz:2023pbu,Kotz:2025lio} we selected and combined the representative pion fits, out of about 100 fits with different functional forms, with the help of the \textsf{mp4lhc} package for \textsf{METAPDF} generation~\cite{Gao:2013bia}. As its name reflects, the package was used for the purpose of the PDF4LHC combination, among other things. Through this procedure, we produced an NLO ensemble \textsf{Fanto10} of Hessian eigenvector sets that quantifies the sum of experimental (aleatoric), parametrization (epistemic), and (in the most recent version in \cite{Kotz:2025lio}) nuclear PDF uncertainties.
The full \textsf{Fanto10} PDF combination follows a sequence of procedures A and B outlines below. 

\subsection{Procedure A. A PDF combination with aleatoric and epistemic uncertainties.}
Given the relative paucity of the available pion data, we adapted a simplified procedure for the PDF model combination that accounts for considerations based on the goodness-of-fit criteria described above. 
\begin{enumerate}
\item For each PDF model $\mathtt{M}$ distinguished by its parametrizations and prior constraints, out of a collection $\mathtt{C}_\mathtt{M}$ of such explored models, we find the best fit corresponding to the minimum of $\chi^2=\chi^2_{\rm LL}+\chi^2_{\rm prior}$ given in Eq.~(\ref{eq:objective}). We keep the models with the likelihood $\chi^2_{\rm LL}(\mathtt{M})$ satisfying
\begin{equation}
    \chi^2_{\rm LL}(\mathtt{M}) \leq \min_{\{\mathtt{M} \in \mathtt{C_M}\}} \chi^2_{\rm LL} + \sqrt{2(N_{pts}-N_{par}(\mathtt{M}))},
    \label{eq:chi2LLM}
\end{equation}
where $\min_{\{\mathtt{M} \in \mathtt{C_M}\}} \chi^2_{\rm LL}$ is the lowest log-likelihood $P(D|\mathtt{M})$ for the fitted pion data $D$ among all tried PDF models $\{\mathtt{M}\}$. $\sqrt{2(N_{pts}-N_{par})}$ is one standard deviation on $\chi^2_{\rm LL}$ with $N_{pts}-N_{par}$ degrees of freedom. We note that this selection does consider the prior model penalties in $\chi^2_{\rm prior}$, viewed as a part of the parametrization. 
\item 
Among the select best fits with low $\chi^2$,  not all result in distinct PDF behaviors. We retain a small number, $\widetilde N$,  of them that produce distinct PDF shapes, which have been selected by ad-hoc criteria in the current implementation. 
\item For each of $\widetilde N$ retained models, we generate a Hessian uncertainty $\Delta_{\rm LL} f_a(x, Q_0; \mathtt{M})$ according to the textbook criterion $\Delta \chi^2_{\rm LL}=1$ at 68\% CL. For the pion fit, we choose $\widetilde N=5$.  
\item We combine the Hessian error bands obtained in the previous step into a single uncertainty using the \textsf{METAPDF} procedure \cite{Gao:2013bia} consisting of the following. First, we generate 100 MC replicas from each error band according to the algorithm 
developed in Ref.~\cite{Hou:2016sho}. The $100\cdot \widetilde N$ replicas are then compressed into a Hessian PDF set using the \textsf{mp4lhc} code, version 2.0. As a part of the pion study, we upgraded this code to support the \textsf{METAPDF} combination for arbitrary hadrons and (anti)quark flavor contents.
\end{enumerate}

As we stated in Ref.~\cite{Kotz:2025lio},
probabilistically, the final PDF set at each value of $(x, Q^2)$ corresponds  to the sum of the probability of each of the ${\tilde N}$ distributions. Those are independent and have identical statistical properties\footnote{The \Fantomas methodology, along with the combination of sets, ensures that the models are applied independently, and resampling is done independently for each model's output. The \textsf{METAPDF} procedure, as resampling method, ensures that the distributions are identically generated for all models' outputs. }. Also, the ${\tilde N}$ models are exchangeable. Hence, the  final \FantoPDF set can be interpreted as coming from the distribution of a 
latent variable, {\it i.e.}, a prior, that captures the underlying influence on the ${\tilde N}$ models -- an underlying truth for PDF shapes~\cite{definetti1937foresight}. This interpretation casts the \Fantomas methodology, incremented with the combination \`a la \textsf{METAPDF}, as effectively capturing the epistemic uncertainty. 

\begin{figure}[t]
    \centering
    \includegraphics[width=0.45\linewidth]{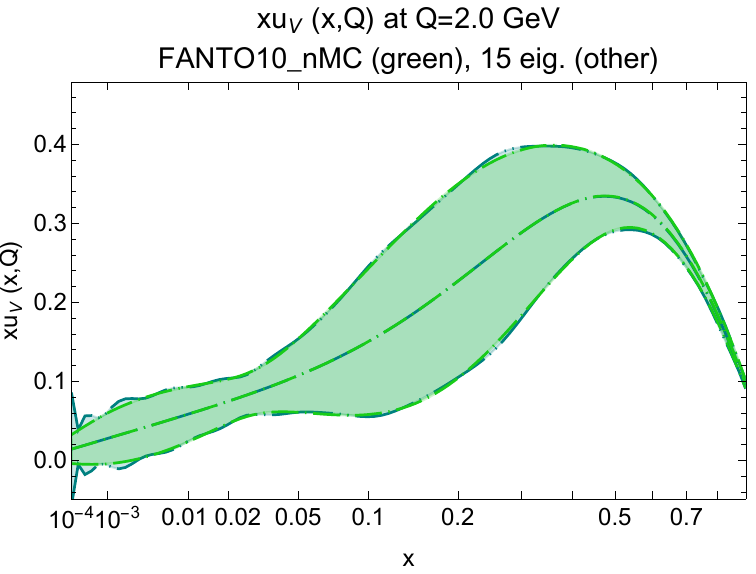}
    \includegraphics[width=0.45\linewidth]{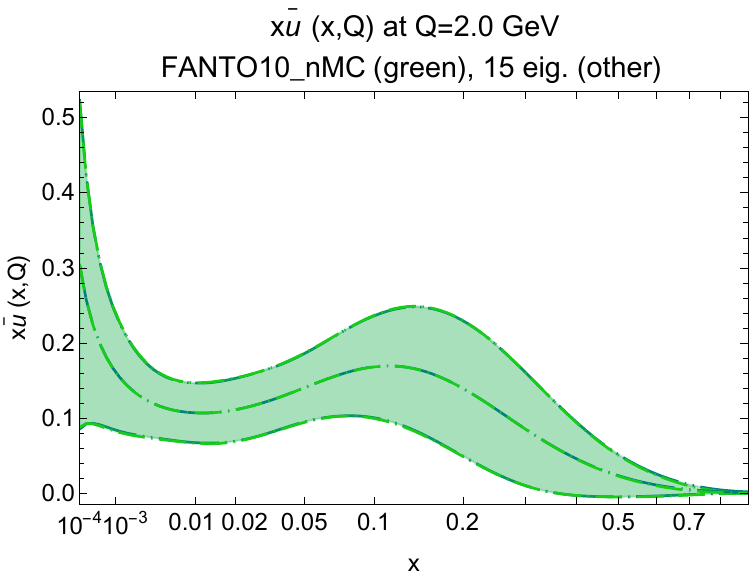}
    \includegraphics[width=0.45\linewidth]{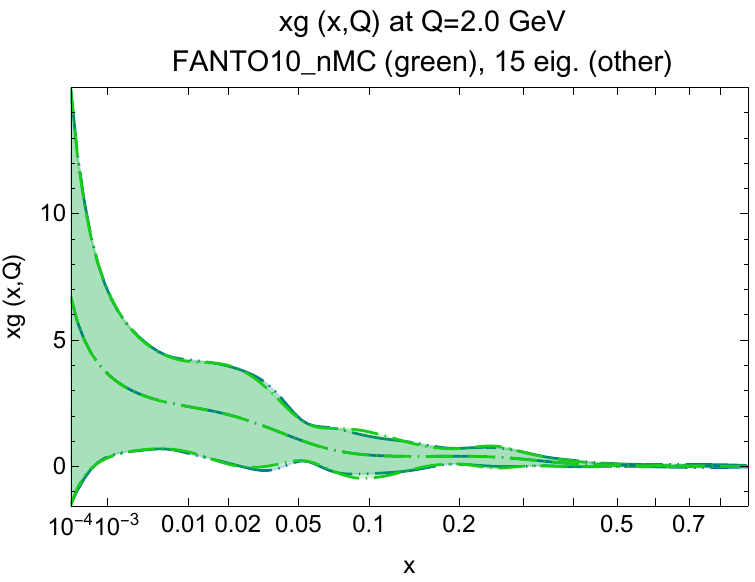}
    \caption{ The aleatoric+epistemic+nuclear uncertainties of the published combination of \textsf{Fanto10\_n15} NLO pion PDFs~\cite{Kotz:2025lio} for $xu_v=xV/2$, $x\bar u=xS/6$, and $xg$ at QCD scale $Q=2$ GeV. Light green curves and filled bands represent the average PDFs and uncertainties of the Monte Carlo replica sample at 68$\%$ C.L. The darker ones stand for the central PDFs and uncertainty bands obtained by the conversion of the above MC replica set into a Hessian ensemble with 15 eigenvector sets.}
    \label{fig:FantoPI}
\end{figure}

\subsection{Procedure B. Including  uncertainties from other nonperturbative functions\label{sec:Nuclear}}
The \Fantomas framework can include other types of uncertainties in addition to the ones discussed above. Ref.~\cite{Kotz:2025lio} describes the combination procedure that includes a third-party input with uncertainties, in that particular case, the nCTEQ15 nuclear PDFs \cite{Kovarik:2015cma} required for the computation of cross sections in pion-nuclear Drell-Yan pair production. The nCTEQ15 Hessian PDFs were converted into Monte-Carlo replicas, then the respective variations in the central pion PDFs were absorbed into the final pion+nuclear uncertainty of the respective \textsf{FantoPDF} ensemble using the \textsf{METAPDF} combination. 
The resulting updated combination for the \texttt{Fanto10} PDFs~\cite{Kotz:2023pbu} are shown in Fig.~\ref{fig:FantoPI}. This figure emphasizes that the final Hessian ensemble approximates well the error bands obtained with intermediate Monte-Carlo replicas, which capture both the variability of the selected central fits due to the \Fantomas functional forms, as well as the error bands around each central fit due to the experimental uncertainty.
This procedure could be followed for including other uncertainties involving third-party inputs, such as those from fragmentation functions in semi-inclusive DIS.

\section{Guidelines}
\label{sec:guidelines}

In the previous Sections, we have illustrated the power of the \Fantomas methodology. In this Section, we will summarize 
the ``dos and don'ts" to follow.\\

\noindent {\bf Dos:}
\begin{enumerate}
    \item \emph{Start with low-degree polynomials}:
Begin by using bare carriers or low-degree polynomials with fixed control points near the endpoints, as outlined in Sec.~\ref{sec:0thfit}. 
\item \emph{Place fixed control points} at the lower and upper ends of the $x$ region probed by the experimental data. Outside of this region, now delineated by the variables $x_{min}$ and $x_{max}$ that were set by placing these CPs, the metamorph will follow the asymptotic behaviors at $x\to 0$ or  $x\to 1$ that are fully controlled by the carrier.
  \item \emph{Add control points incrementally}:
Once a preliminary fit with such basic parametrizations has converged, add one or few NEW CPs according to the sequential approach outlined in Section~\ref{sec:0thfit} to further improve description of data while maintaining the stable convergence and optimizing fitting time. According to the unisolvence theorem, varying positions of FREE CPs for a fixed $N_m$ does not modify the eventual best-fit \meta, but it may affect the rate of convergence. Varying numbers and positions of FIXED CPs and the modulator's stretching power $\alpha_x$ (or possibly the whole stretching function $y(x)$) does change the best-fit \meta. Varying $\alpha_x$ is highly recommended to identify a Bernstein basis that adequately captures the $x$ dependence of data. 
    \item \emph{Set penalties thoughtfully}:
Implement prior constraints with the \texttt{metamorphCollection:Chi2prior()} method and save the files \texttt{steering\_fantomas\_out.txt} and \texttt{metamorphCollectionPrior.cc}  in the \texttt{output/} directory, retaining the \meta and prior settings for the lowest achieved $\chi^2$. The re\-sul\-ting solutions may strongly depend on the prior. 
\end{enumerate}

\noindent {\bf Don'ts:}
\begin{enumerate}
    \item \emph{Avoid overcomplicated polynomial degrees}:
Smooth functions of $x\in [0,1]$ rarely require excessively high polynomial degrees. Although over-fitting is not inevitable, it is possible. Maintain a minimal polynomial degree while ensuring integrability, cf. Sec.~\ref{sec:penal}, and proper endpoint behavior, as discussed in the ``Dos." For example, in the pion PDF fit, a maximum polynomial degree of $N_m=3$ was used, beyond which the data struggled to accommodate for additional flexibility/parameters.
    \item \emph{Beware of oscillatory patterns}:
    One-dimensional Bézier curves are subject to the Runge phenomenon %
    —numerical oscillations arising from equally spaced interpolation points. Mitigate these by:
    \begin{itemize}
        \item Positioning control points unevenly and at sufficient distance from each other.
        \item Adjusting the stretching power $\alpha_x$ to optimize data coverage. \ref{sec:num} provides a detailed analysis of numerical artifacts.
    \end{itemize}

    \item \emph{Monitor condition numbers}:
    Coefficients for Bézier polynomials are computed through inversions of Vandermonde matrices, as shown in equation \ref{eq:Cmatrix}. Numerical stability of solutions is gauged by the condition number. \Fantomas evaluates it in \texttt{metamorph.cc}, as
 described in~\ref{sec:num}, and reports as an error message if the condition number is excessive. 
\begin{python}
  //check the condition number for T                                                    
  double condnum=MetaVector[NMeta].GetConditionNumber();                                
  if (condnum > 10000){                                                                 
    cerr << "WARNING: a high condition number ="<<condnum                               
         << "in metamorph " << MetaVector[NMeta].ID << endl;                            
    cerr << "Check x spacing of its control points" << endl;                            
  } 
\end{python}
Although there is no strict threshold for the condition number, selecting solutions with relatively lower values is advisable to reduce numerical fluctuations and, collaterally, oscillatory effects~\cite{Gautschi1978}. %

\end{enumerate}

\section{Conclusions}
\label{sec:conclusions}

In global QCD analyses of hadron scattering data, inference of  functional forms for nonperturbative functions necessitates comprehensive techniques to systematically span the solution space and quantify associated uncertainties. 
Neural networks are now commonly used for this purpose, 
in parallel with the traditional approaches based on analytical parametrizations, 
but \Bezier curves offer a new effective alternative for solving this complex inverse problem. \Bezier curves rely on {\it control points} that determine the shape of the curve and can be algorithmically chosen either by a human or intelligent program. Variation of the control points supplants direct variations of polynomial coefficients and is computationally efficient. We introduced a C++ toolkit \Fantomas for generation of PDF parametrizations in the \Bezier representation and control of prior constraints on the PDFs. \Fantomas can be deployed together with the program \textsf{mp4lhc} to create combined uncertainty bands using a variety of \Bezier functional forms, or metamorphs, according to the \textsf{METAPDF} PDF combination methodology \cite{Gao:2013bia}. This toolkit has been applied to the determination of pion PDFs~\cite{Kotz:2023pbu}, demonstrating its robustness when implemented inside the \xFitter minimization framework. 

Section~\ref{sec:func_form} of the article sets up the mathematics of the \Bezier polynomial approximants on which the \Fantomas methodology is based. Then, Sec.~\ref{sec:fantomas} reviewed the components of the \Fantomas toolkit and introduced its two realizations as a standalone program or as a module in \xFitter. The former was explained in detail with code snippets in Sec.~\ref{sec:standalone}, and the latter in Sec.~\ref{sec:xFitterImplementation}, together with a short description of the C wrapper to simplify interfacing with with \texttt{Fortran} in Sec.~\ref{sec:CWrapper}. 

Section~\ref{sec:FantoPDFs} focused on a series of simple fits of pion PDFs demonstrating how to deploy the \Fantomas toolkit in practice. We discuss how to progressively increase the flexibility of the functional forms by adding control points and to mitigate non-convergence or overfitting by adding prior penalties using a built-in method of the \Fantomas framework, called \texttt{Chi2prior()}.
 Overfitting arises when a model becomes too complex relative to the data. Using flexible functional forms along with adaptive methodologies does not inherently lead to this issue, but requires a systematic tracking, here implemented through priors. 

After obtaining converged fits for several sets of \meta functional forms and respective error bands for experimental uncertainties, we can combine them using the \textsf{METAPDF} technique. 
Section~\ref{sec:combination} outlines a procedure for such combination, which includes two main components: Procedure A for the combination of models ({\it i.e.}, functional forms), guided by information criteria and quantitative indicators such as the likelihood-ratio test, penalties due to the number of parameters, and distances between functions; and Procedure B for inclusion of uncorrelated uncertainties from external sources.  
The latter is exemplified by the inclusion of nuclear PDF uncertainties in the \texttt{Fanto10\_n15} NLO error PDFs for the pion presented in a separate publication~\cite{Kotz:2025lio}. %

To guide the users in efficient generation of the \Bezier-based parametrizations,  we have compiled a list of ``dos and don'ts" in Sec.~\ref{sec:guidelines}, drawing on the material in the preceding sections and our own experience with fitting such PDFs. This article includes two appendices: \ref{sec:xfit} with an overview of the \xFitter network, and \ref{sec:num} dedicated to the control of numerical artifacts with polynomials of elevated degree. We conclude that the polynomial approximators are no more prone to numerical instabilities, biases, or overfitting than neural networks are. The \Bezier polynomials can approximate smooth 1-dimensional functions to arbitrary accuracy, and they allow for transparent, interpretable exploration of the full diversity of PDF shapes. In practice, low-degree polynomials, enhanced by \Bezier curve features such as control points and stretching parameters, are generally sufficient for modeling one-dimensional shapes at $x \in [0,1]$.

This methodology is developed with an eye on the next generation of precision global analyses of unpolarized PDFs. It may prove useful for modeling other types of non-perturbative functions as well, and it goes along with information-theoretical approaches to quantify the epistemic uncertainty. From a Bayesian perspective, \Fantomas facilitates systematic exploration, or sampling, of a wide range of plausible functional forms~\cite{MengXL:2018,Courtoy:2020fex, Courtoy2023From}, {\it i.e.}, marginalizing over the parameter priors~\cite{DelDebbio:PDF4LHC24}. The C++ code deliverables of this project (both the standalone and \xFitter implementations) are available from our \textsf{GitLab} repository.

\section*{Acknowledgments}
We thank the  \xFitter collaboration members for their assistance with the \xFitter program. 
The authors are extremely thankful to Alim Ablat and Yao Fu for beta testing the standalone code.
This study has been financially supported by 
a National Science Foundation AccelNet project, 
by the U.S.\ Department of Energy under Grant No.~DE-SC0010129,
and 
by the  U.S.\ Department of Energy, Office of Science, Office of Nuclear Physics, 
within the framework of the Saturated Glue (SURGE) Topical Theory Collaboration.
AC and MPC were partially  supported by the UNAM Grant No. DGAPA-PAPIIT IN102225.
The work of TJH at Argonne National Laboratory was supported by the U.S.~Department of Energy under contract DE-AC02-06CH11357.
AC and FO are grateful for the hospitality of the BNL EIC Theory Institute where some of this work was performed. 
PMN is grateful for support from Wu-Ki Tung Endowed Chair in particle physics. PMN an AC thank the Institute of Nuclear Theory at University of Washington for the hospitality during their work on a part of this project.

\appendix

\section{Review of the \xFitter framework}
\label{sec:xfit}

\xFitter is an open-source QCD fitting framework designed to assess the impact of new data on
the extracted parton distribution functions (PDFs) and fragmentation functions (FFs). \xFitter
provides a standard interface to a wide variety of tools to facilitate comparative studies and evaluate constraints
on PDFs/FFs. It has been used for 100+ analyses, including many LHC studies.

\begin{wrapfigure}{BTH}{0.5\textwidth}
  \begin{center}
    \includegraphics[width=0.49\textwidth]{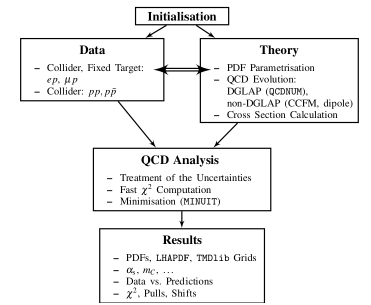}
  \end{center}
  \caption{\xFitter schematic organization. (Fig. from Ref.~\cite{Alekhin:2014irh}) }
  \label{fig:xfit}
\end{wrapfigure}

\xFitter combines experimental data with theoretical calculations to determine the PDFs. The experimental data used in \xFitter come from a wide range of sources, including fixed-target and collider experiments. The specific datasets are selected based on the PDFs being studied and the constraints required to define them.
After selecting the data, \xFitter uses the generated PDFs to compute the theoretical predictions and compare them with the experimental measurements. It optimizes the agreement between the data and theory by modifying the trial PDFs using the \textsf{MINUIT} minimization package from the CERN libraries.

We augmented the \xFitter framework by including the \Fantomas toolkit for advanced PDF parametrizations, and we extended a fit of NLO PDFs in the pion \cite{Novikov:2020snp} available in the \xFitter examples to explore the parametrization dependence with \meta parametric forms \cite{Kotz:2023pbu,Kotz:2025lio}.

The \xFitter fitting algorithm is organized into five stages: 1)~Initialization, 2)~Preparation of data, 3)~Theory calculations, 4)~QCD analysis, and 5)~Output of results. 
This general structure is illustrated in the flowchart of Fig.~\ref{fig:xfit}.
The initialization process declares all necessary variables to be used based on the initial settings. The chosen experimental data sets are compared against the theoretical predictions in the course of the QCD analysis, which calculates and saves the final results at the end.

The main package of \xFitter stores a variety of PDF parameterizations, and it is possible to add a custom parameterization as well. 
We use this option offered by the \xFitter's modular structure to embed the \Fantomas parameterization without interference with the other \xFitter routines, as \Fantomas is (mostly) localized within its own module.

This implementation is a part of the section ``3) Theory" of the flowchart of Fig.~\ref{fig:xfit}. %
Specifically:    
\begin{itemize}  %
\item 
\xFitter implements several parameterizations, and these are contained in the  \texttt{pdfparams/} directory. 
For example, the \texttt{HERAPDF/} directory contains the commonly used HERA PDF parameterization. 
\Fantomas source files are included in \texttt{pdfparams/Fantomas/}, as detailed in Sec.~\ref{sec:xFitterImplementation}. %
\item 
To configure the fit to call the \Fantomas parametrization, we modify \texttt{parameters.yaml} containing the settings of the fit in the following sections: 
    \begin{enumerate}
        \item \textsf{MINUIT:} lists options for the \textsf{MINUIT} fitting routine;
        \item \texttt{Parameters:} list and initial values used for the fitted  parameters passed to \textsf{MINUIT}. The majority of parametrizations store the initial and final parameters in this section. A \texttt{Fantomas} fit with its distinctive format of parameterizations, on the other hand, requires separate steering cards, \texttt{steering\_fantomas.txt} and \texttt{steering\_fantomas\_out.txt} to store the initial and current best-fit \meta parameters and settings, respectively. The \texttt{Parameters} section of \texttt{parameters.yaml} in this case has a simpler fixed format that stores only the initial nil values for deltas, the changes of the \meta parameters, as detailed in Secs.~\ref{sec:xFitterImplementation} and \ref{sec:0thfit};
        \item \texttt{Parametrizations:} call the classes for the chosen parametrization per flavor and lists the parameters to be minimized per flavor. This can be a mix of parametrizations from \Fantomas and other parametrization forms, such as \texttt{HERAPDF};
    \item \texttt{Decompositions:} The \texttt{parameters.yaml} file also specifies the flavor decomposition for chosen hadron. The available decompositions can be found in \texttt{pdfdecomps/}.  For the pion fit, either the \texttt{SU3\_Pion}, for $\pi^-$, or the \texttt{SU3\_PositivePion}, for $\pi^+$, can be chosen if the sea normalization $A_S$ is fitted as a free parameter. If the gluon normalization $A_g$ is fitted instead, the respective decompositions are 
    \texttt{SU3\_Pion\_gluon} and \texttt{SU3\_PositivePion\_gluon};
    \item \texttt{Evolutions:} specifies the external evolution code, {\it e.g.}, QCDNUM, or the use of LHAPDF grid, as well as the class for all involved hadrons and their respective PDFs. This part of the file also specifies the choice of QCD parameters and alike, and it calls the file \texttt{constants.yaml}; \texttt{DefaultEvolution:} sets parameters of the DGLAP evolution; 
    \item \texttt{byReaction:} specifies heavy-quark factorization schemes and relevant QCD parameters by the reaction. For example, at NLO \xFitter 
    provides DIS cross sections in the Thorne-Roberts (TR’)~\cite{Thorne:2006qt} 
    and ACOT~\cite{Aivazis:1993pi,Stavreva:2012bs} heavy-quark schemes.
    \item \texttt{WriteLHAPDF6:} sets the header for the LHAPDF grids.
    \end{enumerate}

\item 
Within \xFitter, the  \texttt{steering.txt} file specifies which data sets are used.

\item The \texttt{constants.yaml} file lists all EW parameters as well as quark masses.

\end{itemize}  
To recap, the \Fantomas extension retains the general structure of the \xFitter framework. 
Thus, it provides access to the full range of \xFitter features, including evaluating functions to compute PDF uncertainties, data pulls, 
$\chi^2$ values for individual data sets and fitting parameters, as well as to export the results into LHAPDF grids and \texttt{xfitter-draw} plotter program. 
Additional information on the \xFitter program can be found in Refs.~\cite{Alekhin:2014irh,xFitterwebsite}.

\section{Numerical artifacts}
\label{sec:num}

Polynomial interpolation is not free of numerical artifacts, one of which is the wiggly or even oscillatory behavior that the interpolating polynomial exhibits under some conditions. When working with flexible polynomials—either by increasing their degree or by the choice and placement of interpolating points—concerns about the smoothness and stability of the interpolation naturally arise. 

\begin{figure*}[t]
     \centering
     \includegraphics[width=0.475\linewidth]{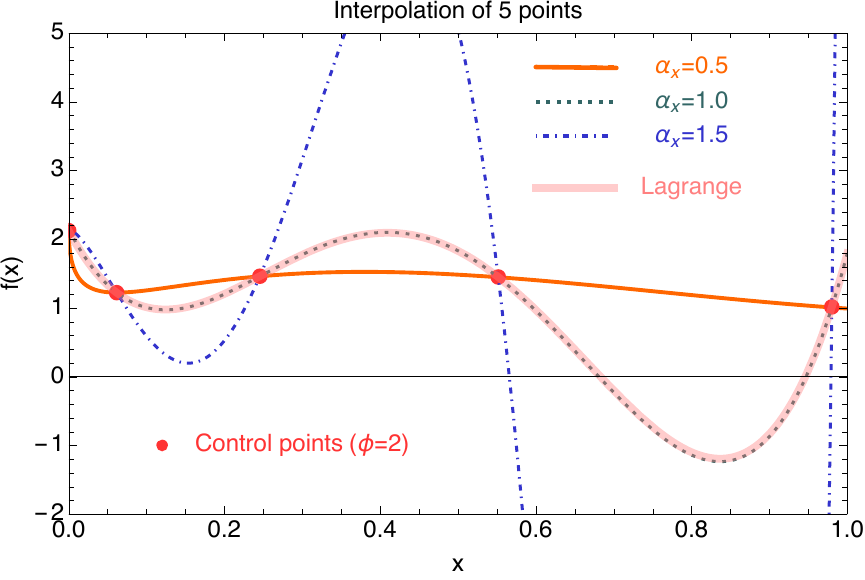}\quad
     \includegraphics[width=0.475\linewidth]{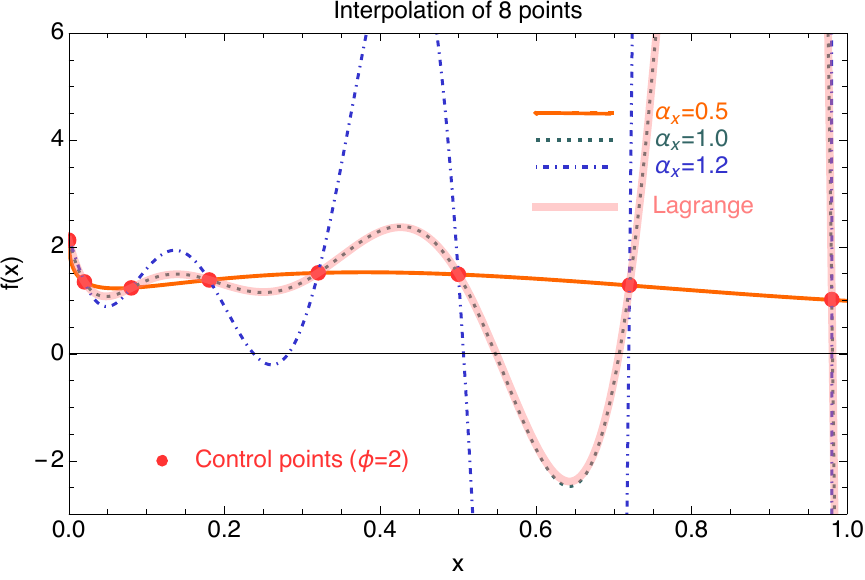}
\caption{Effect of the stretching power $\alpha_x=\mathtt{xPower}$ on the shape of the interpolated function and the oscillatory pattern, for 5 points ({\it l.h.s.}) and 8 points ({\it r.h.s.}) distributed following Eq.~(\ref{eq:phi_dist}). The \Bezier curves obtained with $\alpha_x=1$ correspond to the Lagrange interpolation.
}
\label{fig:PDF-like_oscill}
\end{figure*}

In the context of Bézier curves, these interpolating points are related to the positions of what, in this manuscript, we call {\it the control points}. In Section~\ref{sec:metamorph}, we introduced the mathematical formalism of Bézier curves via control points and illustrated their behavior in Section~\ref{sec:FantoPDFs}. Here, we extend the discussion to analyze how the numerical behavior of the interpolation is affected by the positioning of these control points in our \meta formalism. To do so, we will analyze two \meta %
settings in interpolation mode, which differ by their polynomial degree and the stretching parameter, $\alpha_x$, illustrated in Fig.~\ref{fig:PDF-like_oscill}. We work with square systems only.

\subsection{Parametric vs. 1D \Bezier curves}

\Bezier curves are full two-dimensional parametric curves  --${\cal B}(t)=(x(t),y(t)),\, t\in[0,1]$-- that are defined through a convex hull determined by  {\it external} control points~\cite{pomax}. The control points that define the Bézier hull are connected via de Casteljau’s algorithm to the interpolating points used in our formalism (and which are referred to as {\it control points} in the manuscript), as detailed in Section~\ref{sec:metamorph}.

The \meta formalism, however, is not based on the full two-dimensional parametric \Bezier curve but rather on its $y$-projection, denoted as ${\cal B}(x)$ with $x$ the coordinate~\cite{Elserthesis:2025}, with $N_m+1$ interpolation points $\{x_i, {\cal B}^{(N_m)}(y(x_i))\}, \, i=1, \cdots, N_m+1$, effectively reducing to a one-dimensional interpolation, and is thus subject to numerical artifacts similar to those observed in other polynomial interpolation schemes.

In what follows, we will continue to refer to these interpolation points as control points (CPs), consistent with the terminology used throughout the manuscript. Note that, in contrast to classical parametric \Bezier curves, the resulting \meta function passes exactly through its CPs.

\subsection{Condition Number and Matrix Inversion in Polynomial Interpolation}

Multiple criteria were tested to determine the optimal values of \Fantomas' parameters in order to produce acceptable solutions, particularly to avoid undesired wiggliness.
Oscillatory patterns 
are closely tied to the behavior of the transformation matrix, which in our case is the product of the $T$ and the $M$ matrices. 
The $T$-matrix naturally arises in polynomial interpolations; its entries are successive powers of the control points. Matrices with this structure are called {\it Vandermonde matrices}, whose stability and conditioning have been extensively studied~\cite{Gautschi1962, Gautschi1974, Gautschi1978}.
The product ${\underline{\underline T}}\, \cdot {\underline{\underline M}}$ forms the coefficient matrix of the linear system that must be solved to fully determine the interpolating polynomial. The stability of the interpolation is determined by this composed matrix.
The stability of these fits can be assessed by the condition number %
\begin{eqnarray}
    \kappa\left(\underline{\underline A}\right) = \lVert \underline{\underline A} \rVert\, \lVert \underline{\underline A}^{-1} \rVert\, ,
    \label{eq:cond_numb}
\end{eqnarray}
with  $\lVert \underline{\underline A} \rVert = \sum_{i=1}^{n} \lVert {\underline A}_{i} \rVert_2 = \sum_{i=1}^{n} \left( \sum_{j=1}^{m} |A_{j,i}|^2 \right)^{1/2}$. 
This quantity measures the amplification of input perturbations in the output,  reflecting whether the matrix 
${\underline{\underline T}}\, \cdot {\underline{\underline M}}$ is \textit{ill} or \textit{well}-conditioned. 
Matrices with large condition numbers are prone to substantial output variations from small input perturbations, leading to numerical instability. Consequently, large condition numbers are undesirable, yet they do not always correlate with  oscillatory patterns.
\begingroup
\setlength{\tabcolsep}{10pt} %
\renewcommand{\arraystretch}{1.5} %
    \begin{table}
    \centering
    \begin{tabular}{|c|c|c|c|}
        \hline
        Points & $\alpha_x=0.5$ & $\alpha_x=1.0$ & $\alpha_x=1.5/1.2$\\
        \hline
        5 & 12.2109 & 42.9459 & 479.995\\
        \hline
        8 & 192.45 & 2472.02 & 19683.3\\
        \hline
    \end{tabular}
    \caption{Condition number for the interpolating polynomials of Fig.~\ref{fig:PDF-like_oscill}.}
    \label{tab:conditionnumber}
     \end{table}
\endgroup
Optimizing the condition number becomes a delicate balance between choosing the number of control points and their positions, which is a hallmark of the Runge phenomenon that will be discussed hereafter. Table~\ref{tab:conditionnumber} provides the condition numbers for the six interpolating functions illustrated in Fig.~\ref{fig:PDF-like_oscill}. We observe that the condition number is larger for curves with more pronounced oscillations. However, this comparison is relative for each setting, {\it e.g.}, for each value of $N_m$. Because $\underline{\underline{M}}$ has a fixed form once $N_m$  is set, optimizing the condition number relies solely on how $\underline{\underline{T}}$ is configured~\cite{vdm26}. %

Moreover, the inversion of the matrix $\underline{\underline{T}}$ is computationally costly for high values of 
$N_m$. %
We adopted the LU-decomposition method, implemented in a separate header file, LUPinverse.h. This file contains a function for inverse calculation, which is called in the main code wherever matrix inversion is needed. The inversion time for various $N_m+1$ points are given in Table~\ref{tab:inversion_time}.
    \begin{table}[t]
    \centering
    \begin{tabular}{|c|c|}
        \hline
        Points &   LU Decomposition\\
        \hline
        10 &  0.0007s\\
        \hline
        15 &  0.085s\\
        \hline
        20  & 2.221s\\
        \hline
        25  & 1m11.723s\\
        \hline
        30 & 38m\\
        \hline
    \end{tabular}
    \caption{Time required for $\underline{\underline{T}}$ matrix inversion with the LU decomposition method.}
    \label{tab:inversion_time}
     \end{table}

\subsection{Runge Phenomenon in Polynomial Interpolation}

A classical example of interpolation instability is the Runge phenomenon, characterized by large variations in the derivatives of an interpolating function near its end-points when evaluated at equidistantly spaced or evenly distributed nodes. 
For squared systems like the ones considered in Section~\ref{sec:metamorph}, 
two equivalent closed-form solutions to the interpolation problem exist--  one is based on  Lagrange polynomials and the other on generalized Bernstein polynomials (see Ref.~\cite{Courtoy:2020fex} for details).

The Runge phenomenon is generally illustrated using the Lagrange polynomial-based interpolation. The full 2D parametric Bézier curves  are known to avoid such spurious effects, while it is not the case for the 1D \meta. The latter  share characteristics with Lagrange polynomial interpolation.
Nonetheless, the flexibility in the choice of control points and scaling of the argument of the Bernstein polynomial, through $\alpha_x$ in Eq.~(\ref{eq:stretch}), allows some mitigation of these Runge artifacts.
In Fig.~\ref{fig:PDF-like_oscill}, we illustrate the behavior of interpolating curves through our formalism for 5 and 8 evenly distributed control points, following
\begin{equation}
    x_i = \left( x_{min} + \frac{x_{max} - x_{min}}{N_m} \, i \right)^\phi, \, \mbox{ for}\quad i=0,\ldots, N_m.
    \label{eq:phi_dist}
\end{equation}
with $\phi=2$, are chosen from a known reference or ``truth" function, and then these are interpolated for various values of $\alpha_x$. 
The case $\alpha_x=1$ corresponds to the standard Lagrange polynomial interpolation and exhibits oscillations for $N_m$ a low as 4 (left panel). By tweaking $\alpha_x$, we reduce the wiggly pattern, a reduction also reflected in improved condition numbers. The optimal choice to minimize the appearance of high derivatives is $(\phi=2, \alpha_x=0.5)$.\\

For practical use of the \Fantomas framework, caution is advised when increasing the degree of the polynomial substantially, especially when evenly spaced control points are chosen. For this study, we have explored the possibility of oscillatory solutions, and found them difficult to induce. Notably, no such patterns were observed in the first physics application of the \Fantomas framework, {\it i.e.}, the analysis of the pion PDFs~\cite{Kotz:2023pbu} where the highest degree setting was $N_m=3$ with unevenly spaced control points. It is important to note that oscillations, when they occur, are not necessarily spurious numerical artifacts.
\section{A Mathematica implementation}
\label{sec:mathematica}

 Our methodology is best illustrated through simple interpolation examples, {\it e.g.}, as proposed in Ref.~\cite{Courtoy:2020fex} to study the large-$x$ behavior of proton PDFs. For that purpose, Fig.~\ref{fig:Mathematica} provides a \textsf{Wolfram Mathematica} implementation of the \Bezier polynomials that are a part of \meta modulators. The $\mathtt{Bezier}$ module can easily be extended to include minimization instead of interpolation. Here, it is shown for $x\, f(x)=F_a^{\rm car}(x)\times {\cal B}^{(N_m)}(x^{\alpha_x})$, and can easily be adapted to other implementations.
\begin{figure}
\begin{python}
  Fc[x_, {a0_, a1_, a2_}] := a0*x^a1 (1 - x)^a2;
  
  Bezier[xPower_ : 1, k_ : 1, n_ : 1, q0_ : 1.3] := 
  Module[{m, nlist, M, t, T, p, l, x, P, C, f, b, B, pts, plotmod, 
   errplotmod},
  Params = List[];
  AppendTo[Params, xPower];
  AppendTo[Params, k];
  AppendTo[Params, n];
  AppendTo[Params, q0];
  nlist = Range[0, n];
  m[p_, l_] := (-1)^(p - l) Binomial[n, l] Binomial[n - l, n - p];
  M = Array[m, {n + 1, n + 1}, {{0, n}, {0, n}}];
  t[x_, p_] := x^(p*xPower);
  T = ArrayReshape[
    Table[t[x, p], {x, xlist}, {p, nlist}], {k + 1, n + 1}];
  P = ArrayReshape[pdflist/Fc[xlist, auV], {k + 1, 1}];
  C = Inverse[M] . Inverse[Rationalize[Transpose[T] . T, 0]] . 
    Transpose[T] . P;
  coef = C;
  ]
  
  FBezier[x_, xPower_, n_, coef_] := 
  Sum[coef[[nu + 1, 1]]*
    Binomial[n, nu] x^(nu*xPower) (1 - x^xPower)^(n - nu), {nu, 0, n}];
\end{python}
\caption{A Mathematica implementation of the \Bezier polynomial. \label{fig:Mathematica}}
\end{figure}
\bibliographystyle{elsarticle-num}
\bibliography{extra,main}

\end{document}